\documentclass[prd,twocolumn,superscriptaddress,nofootinbib,noshowpacs,preprintnumbers,longbibliography,floatfix]{revtex4-1}
\usepackage[utf8]{inputenc}
\usepackage{graphicx}
\usepackage{float}
\usepackage{amssymb}
\usepackage{amsmath}  
\usepackage{mathtools}
\usepackage{dsfont}
\usepackage{array}
\usepackage{bm,fixmath}
\usepackage{subfigure}
\usepackage{mathrsfs}
\usepackage{pifont}
\usepackage{multirow}
\usepackage{upgreek}
\usepackage{xcolor}
\usepackage{bm}
\usepackage{bbm}
\usepackage{physics}
\usepackage{comment}
\usepackage{appendix}
\DeclarePairedDelimiter\set\{\}
\usepackage[pdftex,
            pdftitle={Topological terms in (3+1)D},
            pdfauthor={Angus Kan, Lena Funcke, Stefan Kuehn, Luca Dellantonio, Jan F. Haase, Jinglei Zhang, Christine A. Muschik, Karl Jansen},
            bookmarks,
            colorlinks,
            linkcolor=myblue,
            citecolor=mymagenta,
            menucolor=black,
            urlcolor=myblue,
            plainpages=false,
            pdfpagelabels,
            hypertexnames=false]{hyperref}
\usepackage{verbatim}
\usepackage{slashed}
\usepackage{cleveref}

\definecolor{mymagenta}{RGB}{200, 0, 100}
\definecolor{myblue}{RGB}{45, 48, 146}

\definecolor{mygreen}{RGB}{0, 126, 0}

\newcommand{\Ol}{\ensuremath{\mathcal{O}}}

\makeatletter
\def\mathcolor#1#{\@mathcolor{#1}}
\def\@mathcolor#1#2#3{%
  \protect\leavevmode
  \begingroup
    \color#1{#2}#3%
  \endgroup
}
\makeatother

\begin{document}
\title{Investigating a (3+1)D Topological \texorpdfstring{\boldmath{$\theta$}-}{}Term in the Hamiltonian Formulation of Lattice Gauge Theories for Quantum and Classical Simulations}

\author{Angus Kan}
\thanks{These two authors contributed equally.}
\affiliation{Institute for Quantum Computing and Department of Physics \& Astronomy, University of Waterloo, Waterloo, Ontario, N2L 3G1, Canada}
\author{Lena Funcke}
\thanks{These two authors contributed equally.}
\affiliation{Perimeter Institute for Theoretical Physics, 31 Caroline Street North, Waterloo, Ontario, N2L 2Y5, Canada}
\author{Stefan Kühn}
\affiliation{Computation-Based Science and Technology Research Center, The Cyprus Institute, 20 Kavafi Street, 2121 Nicosia, Cyprus}
\author{Luca Dellantonio}
\affiliation{Institute for Quantum Computing and Department of Physics \& Astronomy, University of Waterloo, Waterloo, Ontario, N2L 3G1, Canada}
\author{\mbox{Jinglei Zhang}}
\affiliation{Institute for Quantum Computing and Department of Physics \& Astronomy, University of Waterloo, Waterloo, Ontario, N2L 3G1, Canada}
\author{\mbox{Jan F. Haase}}
\affiliation{Institute for Quantum Computing and Department of Physics \& Astronomy, University of Waterloo, Waterloo, Ontario, N2L 3G1, Canada}
\affiliation{Institute of Theoretical Physics and IQST, Universität Ulm, Albert-Einstein-Allee 11, D-89069 Ulm, Germany}
\author{Christine A. Muschik}
\affiliation{Institute for Quantum Computing and Department of Physics \& Astronomy, University of Waterloo, Waterloo, Ontario, N2L 3G1, Canada}
\affiliation{Perimeter Institute for Theoretical Physics, 31 Caroline Street North, Waterloo, Ontario, N2L 2Y5, Canada}
\author{Karl Jansen}
\affiliation{NIC, DESY, Platanenallee 6, D-15738 Zeuthen, Germany}

\date{\today}

\begin{abstract}
Quantum technologies offer the prospect to efficiently simulate sign-problem afflicted regimes in lattice field theory, such as the presence of topological terms, chemical potentials, and out-of-equilibrium dynamics. In this work, we derive the (3+1)D topological $\theta$-term for Abelian and non-Abelian lattice gauge theories in the Hamiltonian formulation, paving the way towards Hamiltonian-based simulations of such terms on quantum and classical computers. We further study numerically the zero-temperature phase structure of a (3+1)D U(1) lattice gauge theory with the $\theta$-term via exact diagonalization for a single periodic cube. In the strong coupling regime, our results suggest the occurrence of a phase transition at constant values of $\theta$, as indicated by an avoided level-crossing and abrupt changes in the plaquette expectation value, the electric energy density, and the topological charge density. These results could in principle be cross-checked  by the recently developed (3+1)D tensor network methods and quantum simulations, once sufficient resources become available.
\end{abstract}

\maketitle

\section{Introduction}
\label{sec:intro}

Numerical simulations based on Markov chain Monte Carlo (MCMC) methods of lattice gauge theories have had unprecedented success in computing various non-perturbative aspects of fundamental particle interactions~\cite{Aoki:2019cca}. However, in certain parameter regimes they face a major obstacle in the sign problem that prohibits the simulations of, e.g., topological terms and chemical potentials~\cite{Troyer:2004ge,Gattringer:2016kco}. In addition, MCMC methods rely on formulating the theory in Euclidean spacetime, which prevents a direct simulation of real-time dynamics. One promising approach to overcome these limitations are Tensor Networks (TN), which have successfully been used to simulate lattice gauge theories in (1+1)D (see, e.g., Refs.~\cite{Banuls:2018jag,Banuls2019} and references therein), in (2+1)D~\cite{Kuramashi:2018mmi,Felser2019}, and recently even in (3+1)D~\cite{Magnifico:2020bqt}. In particular, techniques based on TN do not suffer from the sign problem. This allows for exploring regimes intractable with MCMC methods~\cite{Banuls2016a,Silvi2016,Felser2019}, such as topological $\theta$-terms in (1+1)D~\cite{Byrnes:2002gj,Buyens:2017crb,Funcke:2019zna}. Beyond classical algorithms, quantum simulations provide a promising tool to overcome these numerical challenges in the future. First proof-of-concept simulations of Abelian and non-Abelian gauge theories in (1+1)D and in (2+1)D have already been accomplished~\cite{Martinez:2016yna, Kokail:2018eiw,Klco:2018kyo,schweizer2019floquet,gorg2019realization,Mil:2019pbt,klco20202,Yang2020,atas20212,ciavarella2021trailhead,rahman20212}. Furthermore, experimentally realizable schemes for simulating (3+1)D gauge theories have been proposed~\cite{gonzalez2017quantum,bender2018digital,banuls2020simulating}.

With the rapid progress in the development of classical and quantum algorithms in higher dimensions, the simulation of topological terms in (3+1)D is coming into reach. These simulations are particularly relevant for one of the most puzzling aspects of the Standard Model (SM) of particle physics, which is the topological $\theta$-term in quantum chromodynamics (QCD). While this term violates the combined symmetry of charge conjugation and parity (CP) and thus distinguishes matter from antimatter, there is no experimental evidence for CP violation in QCD. Instead, measurements of the neutron electric dipole moment~\cite{Crewther:1979,Abel:2020} constrain the parameter of the $\theta$-term to be smaller than $\sim 10^{-10}$, which results in a fine tuning problem. This so-called strong CP problem has triggered an immense amount of model building beyond the SM (see Ref.~\cite{Hook:2018dlk} for a review) and lattice QCD studies of proposed solutions~(see, e.g., Refs.~\cite{Borsanyi:2016ksw,Dragos:2019oxn,Aoki:2019cca,Alexandrou:2020bkd} and references therein).

While conventional lattice MCMC methods rely on the Lagrangian formulation, the Hamiltonian formalism is particularly suited for both quantum simulations and tensor network methods, and this has recently sparked tremendous interest in expressing lattice field theories in this formulation. Topological terms in (1+1)D~\cite{Coleman:1976uz} and (2+1)D~\cite{Deser:1982vy,Deser:1981wh,Witten:2015aoa} have already been explored in both Lagrangian and Hamiltonian formulations, but the $\theta$-term of (3+1)D lattice gauge theories has so far been treated only in the Lagrangian formulation (see Ref.~\cite{Alexandrou:2017hqw} for a review). In the current paper, we fill this gap by deriving the topological $\theta$-term of (3+1)D lattice gauge theories in the Hamiltonian plaquette formulation, using the transfer matrix method~\cite{creutz1977gauge}. This paves the way towards quantum and tensor network simulations of the topological term of the SM of particle physics.

In our paper, starting with the Lagrangian definition in Ref.~\cite{DiVecchia:1981aev}, we first derive the topological $\theta$-term for generic (3+1)D (non-)Abelian lattice gauge theories in the Hamiltonian formulation, which is given by
\begin{align}
    \theta \hat{Q} &= -\frac{ig^2 \theta}{8\pi^2 a}\sum_{\vec{n},i,j,k,b}\varepsilon_{ijk}\text{Tr}\left[\hat{E}_{\vec{n},i}^b \lambda^b\left(\hat{U}_{\vec{n},jk}-\hat{U}_{\vec{n},jk}^\dag\right)\right].
\end{align}
Here, $\hat{Q}$ is the topological charge, $g$ is the coupling constant, $\theta$ a tunable parameter, $a$ is the lattice spacing, $\vec{n}$ denotes a lattice site, $\varepsilon_{ijk}$ is the 3D Levi-Civita symbol, $\{\lambda^b\}$ is the set of Hermitian gauge group generators for the fundamental representation, and $\hat{E}_{\vec{n},i}^b$ and $\hat{U}_{\vec{n},jk}$ are the electric field and plaquette operators, respectively. 

We then focus on a particular example and perform numerical calculations for a (3+1)D pure compact U($1$) lattice gauge theory in the Hamiltonian formulation using exact diagonalization for a single cube. For gauge couplings $\beta \equiv 1/g^2 \lesssim 0.75$, we find indications of a phase transition at constant values of $\theta$, signaled by spikes in the plaquette expectation value, a jump in both the electric energy density and the topological charge density, and an avoided level-crossing in the ground state energy. 

Our results are particularly relevant in the light of recent and earlier analytical studies of the phase diagram of (3+1)D pure compact U($1$) lattice gauge theory with a $\theta$-term (see, e.g., Refs.~\cite{Cardy:1981qy,cardy1982duality,honda2020topological}). Based on free-energy arguments~\cite{Cardy:1981qy}, duality arguments~\cite{cardy1982duality}, and anomaly matching conditions~\cite{honda2020topological}, it was predicted that a phase transition appears at $\theta=\pi$ and small $\beta$ in the confining phase, which disappears at large $\beta$ in the Coulomb phase. Unlike in pure QCD, where the phase transition at $\theta=\pi$ is known to be of first order~\cite{Dashen1971,DiVecchia:1980yfw,Witten:1980sp}, the order of the transition in the pure compact U($1$) gauge theory is unknown~\cite{Cardy:1981qy,cardy1982duality}. Interestingly, our results indicate that for U(1) this transition is not first order. At the same time, our signs for the phase transition vanish for large $\beta$, which qualitatively agrees with the analytical predictions. Our results for the phase diagram could in principle be cross-checked by (3+1)D tensor network methods~\cite{Magnifico:2020bqt} and quantum simulations once sufficient resources become available. Thus, our work paves the way for classical and quantum computations of the topological $\theta$-term in (3+1)D lattice gauge theories, using the method developed in Ref.~\cite{haase2021resource}. Eventually, this will enable the first mapping of the complete phase diagram and a detailed study of the properties of the obtained phases.

The paper is organized as follows. In Sec.~\ref{sec:lattice}, we review the transfer matrix derivation of the Kogut-Susskind Hamiltonian from Wilson's lattice action. In Sec.~\ref{sec:top}, we derive the topological $\theta$-term using the transfer matrix method. In Sec.~\ref{sec:results}, we provide our numerical results, which indicate a phase transition at constant values of $\theta$. In Sec.~\ref{sec:conclusion}, we summarize and discuss our results, including the prospects for simulating the $\theta$-term using MCMC methods, TN, and quantum simulations. 

Throughout the paper, we disregard the Einstein notation and explicitly display all sums. In order to distinguish the variables in the Lagrangian formulation and operators in the Hamiltonian formulation, we express the latter with a hat ($~\hat{}~$) symbol.

\section{Lattice Formulation}
\label{sec:lattice}

In this section, for the convenience of readers who are unfamiliar with lattice gauge theories, we review two standard approaches to lattice gauge theory, namely Wilson's Lagrangian approach~\cite{wilson1974confinement} and Kogut-Susskind's Hamiltonian approach~\cite{kogut1975hamiltonian}. Moreover, we review the derivation of the Kogut-Susskind Hamiltonian from Wilson's lattice action using the transfer matrix method~\cite{creutz1977gauge}, since it will also be used for the derivation of the topological $\theta$-term. 

In the Lagrangian formulation, the standard approach introduced by Wilson~\cite{wilson1974confinement} defines gauge theories on a hypercubic lattice with spacing $a$ in Euclidean spacetime. The sites are labelled by a four-vector $\vec{n} \equiv \sum_{\mu=0}^{3} n_\mu \hat{\mu}$, where $n_0$ denotes the temporal component, $n_i$ with $i \in \{1,2,3\}$ are the spatial components, and $\hat{\mu}$ is a unit vector in the direction $\mu$. In the Hamiltonian formulation, time is continuous, and thus, gauge theories are defined on a cubic lattice. The links are denoted by their originating sites $\vec{n}$ and directions $\mu$. On the lattice, the vector potential, which starts from site $\vec{n}$ and points to direction $\mu$, is represented by $A_{\vec{n},\mu} = \sum_b A_{\vec{n},\mu}^b \lambda^b$, where $A_{\vec{n},\mu}^b$ are real-valued vector fields that correspond to the generators $\lambda^b$ of the gauge group~\cite{rothe2005lattice}, which become trivial in the case of the U(1) gauge theory. Then, the discretized field strength tensor is defined as
\begin{equation}
    F_{\vec{n},\mu \nu}^b \equiv \frac{1}{a}\left(A_{\vec{n}+\hat{\mu},\nu}^b-A_{\vec{n},\nu}^b- A_{\vec{n}+\hat{\nu},\mu}^b + A_{\vec{n},\mu}^b \right).
\end{equation}
Using these degrees of freedom, one defines a link variable
\begin{equation}
   U_{\vec{n},\mu} \equiv  e^{iga \sum_b A_{\vec{n},\mu}^b \lambda^b}.
\end{equation}
In terms of the link variable, we construct a gauge-invariant plaquette variable
\begin{equation}
    U_{\vec{n},\mu \nu} \equiv U_{\vec{n},\mu}U_{\vec{n}+\hat{\mu},\nu}U_{\vec{n}+\hat{\nu},\mu}^\dagger U_{\vec{n},\nu}^\dagger = e^{iga^2 \sum_b F_{\vec{n},\mu \nu}^b \lambda^b},
    \label{eq:plaqvar}
\end{equation}
which is a product of link variables around a closed loop on the lattice (see the right panel of Fig.~\ref{fig:periodic_cube} for an illustration). Using the fact that $F_{\vec{n},\mu\nu}^b=-F_{\vec{n},\nu\mu}^b$, one can show that $U_{\vec{n},\mu \nu}^\dag = U_{\vec{n},\nu\mu}$. In terms of the plaquette variables, the gauge-invariant Wilson's gauge-field action on the lattice reads~\cite{wilson1974confinement}
\begin{equation}
    S_W = -\frac{1}{2g^2} \sum_{\vec{n}} \sum_{\mu, \nu; \nu>\mu}\text{Tr}\left[U_{\vec{n},\mu \nu} + U_{\vec{n},\mu \nu}^{\dag}\right].
    \label{eq:Wilson}
\end{equation}
For small $a$, we can expand the plaquette variables as an exponential function of $a$, as defined in Eq.~\eqref{eq:plaqvar}, and obtain
\begin{align}
    S_W &= -\frac{1}{2g^2} \sum_{\vec{n}} \sum_{\mu, \nu}\text{Tr}\left[U_{\vec{n},\mu \nu}\right] \nonumber\\
    &\xrightarrow[a\approx 0]{} \frac{a^4}{4} \sum_{\vec{n},\mu,\nu,b}F_{\vec{n},\mu \nu}^b F_{\vec{n},\mu \nu}^b \nonumber\\ &\xrightarrow[a\rightarrow 0]{} \frac{1}{4}\int d^4x \sum_{\mu,\nu,b}F_{\mu\nu}^b(x)F_{\mu\nu}^b(x),
\end{align}
where in the first line, we have used the fact that $U_{\vec{n},\mu \nu}^\dag = U_{\vec{n},\nu\mu}$, and in the second line, the linear term vanishes in the expansion due to $F_{\vec{n},\mu\nu}^b=-F_{\vec{n},\nu\mu}^b$. Thus, only the quadratic term in the action survives. In the continuum limit, where $a\rightarrow 0$, the sum becomes an integral and therefore yields the correct continuum expression.

In the following, we reproduce the derivation of the Kogut-Susskind Hamiltonian using the transfer matrix method~\cite{creutz1977gauge}. The action and the Hamiltonian are related by the partition function, which is defined as
\begin{equation}
    Z = \int DU e^{-S}  = \text{Tr}\left[\left(e^{-a_0 \hat{H}}\right)^N\right],
    \label{eq:partition}
\end{equation}
where $\int DU$ is an integral over the gauge group~\cite{rothe2005lattice}, $a_0$ is the temporal lattice spacing, and $N$ is the number of time steps. Since, in the Hamiltonian formulation, time and space are not treated isotropically, we hereafter denote the temporal and spatial lattice spacing as $a_0$ and $a$, respectively. The transfer matrix is defined as
\begin{equation}
    \hat{T} \equiv e^{-a_0 \hat{H}}.
\end{equation}
The Hamiltonian is defined through the transfer matrix in the temporal continuum limit, where $N\rightarrow \infty$ and $a_0\rightarrow 0$ with $t=N a_0$ fixed. The transfer matrix can be expressed in a complete and orthonormal product basis
\begin{equation}
    \{ \ket{U} = \prod_{\vec{n},i} \ket{U_{\vec{n},i}} \},
    \label{eq:prodbasis}
\end{equation}
where $\ket{U_{\vec{n},i}}$ is an element of the gauge group on the link $(\vec{n},i)$. The inner product and the completeness relation in this basis are given by
\begin{equation}
    \braket{U'}{U} = \prod_{\vec{n},i} \delta(U_{\vec{n},i}',U_{\vec{n},i}), \quad \int DU \ketbra{U}{U}=1.
    \label{eq:basis}
\end{equation}
In this basis, we prove the relation in Eq.~\eqref{eq:partition}
\begin{align}
    Z &= \int DU e^{-S} \nonumber \\
      &=\int DU \bra{U_t} \hat{T} \ket{U_{t-a_0}} \bra{U_{t-a_0}} \hat{T} \ket{U_{t-a_0}} \nonumber \\
      &\quad ...\bra{U_{a_0}} \hat{T} \ket{U_0} \nonumber \\
      &= \int DU \bra{U} \hat{T}^N \ket{U} = \text{Tr}[\hat{T}^N],
      \label{eq:partition2}
\end{align}
where in the second equality, we split the Euclidean path integral $e^{-S}$ into $N$ infinitesimal Euclidean evolution operators, i.e., the transfer matrix, and in the last two equalities, we have used the completeness relation and imposed periodic boundary conditions in the temporal direction such that $\ket{U_t}= \ket{U_0}$.

Working in the temporal gauge $\hat{U}_{\vec{n},0}= 1$, the elements of the transfer matrix, which satisfy Eq.~\eqref{eq:partition2}, are
\begin{align}
\begin{split}
    \bra{U'} \hat{T} \ket{U}& = e^{\frac{a}{2g^2 a_0} \sum_{\vec{n},i} \text{Tr}\left[U_{\vec{n},i}'U_{\vec{n},i}^\dag + U_{\vec{n},i}^{\dag'} U_{\vec{n},i}\right]} \\
    & \quad\times e^{\frac{a_0}{2g^2 a} \sum_{\vec{n},j,k}\text{Tr}\left[U_{\vec{n},jk} + U_{\vec{n},jk}^\dag\right] },
    \end{split}
\end{align}
where $U_{\vec{n},i}'$ and $U_{\vec{n},i}$ are link variables from consecutive time slices. Then, we express the transfer matrix in terms of the matrix operators
\begin{equation}
    \hat{U}_{\vec{n},i} \ket{U} = {U}_{\vec{n},i} \ket{U},
    \label{eq:usefulop1}
\end{equation}
which are diagonal in the product basis $\ket{U}$, and the unitary operators
\begin{equation}
    \hat{R}_{\vec{n},i}(g) \ket{U} = \ket{U'},
\end{equation}
where only $\ket{U_{\vec{n},i}}$ is changed in $\ket{U}$, i.e.,
\begin{gather}
    \ket{U'_{\vec{n},i}} = \ket{g_{\vec{n},i}U_{\vec{n},i}}.
\end{gather}
Here, $g_{\vec{n},i}$ is an element in our unitary group, which is parametrized as
\begin{equation}
    g_{\vec{n},i} = e^{i\sum_b x_{\vec{n},i}^b \lambda^b},
\end{equation}
where $x_{\vec{n},i}^b \in \mathbbm{R}$ are the group parameters, and the group generators for the fundamental representation are normalized such that
\begin{equation}
    \text{Tr}[\lambda^a\lambda^b]=\delta_{ab}.
\end{equation}
Now, the unitary operators can be parametrized as
\begin{equation}
    \hat{R}_{\vec{n},i}(g_{\vec{n},i}) \equiv e^{i \sum_b x_{\vec{n},i}^b \hat{E}^{b}_{\vec{n},i}},
    \label{eq:usefulop2}
\end{equation}
where we have introduced the electric field operators $\hat{E}^{b}_{\vec{n},i}$ that act on the links (see right panel of Fig.~\ref{fig:periodic_cube} for an illustration) and are conjugate to the link operators $\hat{U}_{\vec{n},i}$. They satisfy the commutation relations
\begin{gather}
    \left[\hat{E}^{a}_{\vec{n},i},\hat{E}^{b}_{\vec{n},i}\right] = i\sum_c f^{abc}\hat{E}^{c}_{\vec{n},i} \label{eq:EEcomm}, \\
    \left[\hat{E}^{a}_{\vec{n},i},\hat{U}_{\vec{n},i}\right] = -\lambda^a \hat{U}_{\vec{n},i}\label{eq:EUcomm},
\end{gather}
where $f^{abc}$ are the structure constants of the gauge group. In a U(1) theory, there is only one generator $\lambda^a =1$ for the fundamental representation and one operator $\hat{E}_{\vec{n},i}^a=\hat{E}_{\vec{n},i}$, and the structure constants vanish. Thus, Eq.~\eqref{eq:EEcomm} becomes trivial, and Eq.~\eqref{eq:EUcomm} can be simplified. In terms of these operators, we can write the transfer operator as
\begin{align}
    \hat{T} &= \prod_{\vec{n},i}\int_{g \in G}  dg_{\vec{n},i} \hat{R}_{\vec{n},i}(g_{\vec{n},i})e^{\frac{a}{2g^2 a_0}  \text{Tr}\left[g_{\vec{n},i} + g_{\vec{n},i}^\dag\right]} \nonumber \\
    & \quad \times e^{\frac{a_0}{2g^2 a} \sum_{\vec{n},j,k}\text{Tr}\left[\hat{U}_{\vec{n},jk} + \hat{U}_{\vec{n},jk}^\dag\right] }\nonumber \\
    &= \prod_{\vec{n},i}\int \left(\prod_{b} dx_{\vec{n},i}^b\right) e^{i\sum_b x_{\vec{n},i}^b \hat{E}^{b}_{\vec{n},i} } \nonumber \\
    &\quad \times e^{\frac{a}{2g^2 a_0} \text{Tr}\left[2\cos(\sum_b x_{\vec{n},i}^b \lambda^b)\right]}\nonumber\\
    & \quad \times e^{\frac{a_0}{2g^2 a} \sum_{\vec{n},j,k}\text{Tr}\left[\hat{U}_{\vec{n},jk} + \hat{U}_{\vec{n},jk}^\dag\right] }.
\end{align}
In the continuum limit, as $a_0 \rightarrow 0$, the integral is dominated by the maximum of $\text{Tr}\left[2\cos( \sum_b \lambda^b x_{\vec{n},i}^b)\right]$. Expanding around $x_{\vec{n},i}^b = 0$, where the maximum is located, we have
\begin{equation}
   \text{Tr}\left[2\cos(\sum_b  \lambda^b x_{\vec{n},i}^b)\right] \approx  2D - \sum_b x_{\vec{n},i}^b x_{\vec{n},i}^b,
\end{equation}
where $D$ is the dimension of the group generators. Inserting the expansion into the integral, we obtain a Gaussian integral, which evaluates to
\begin{equation}
    \hat{T} \propto e^{-a_0\left(\frac{g^2}{2a} \sum_{\vec{n},i,b} \hat{E}^b_{\vec{n},i}\hat{E}^b_{\vec{n},i} - \frac{1}{2g^2 a} \sum_{\vec{n},j,k}\text{Tr}\left[\hat{U}_{\vec{n},jk} + \hat{U}_{\vec{n},jk}^\dag\right]\right)}.
    \label{eq:Gaussian}
\end{equation}
From this, we can directly read off the Kogut-Susskind pure gauge Hamiltonian~\cite{kogut1975hamiltonian}
\begin{equation}
    \hat{H}_{KS} = \frac{g^2}{2a} \sum_{\vec{n},i,b} \hat{E}^b_{\vec{n},i}\hat{E}^b_{\vec{n},i} - \frac{1}{2g^2 a} \sum_{\vec{n},j,k}\text{Tr}\left[\hat{U}_{\vec{n},jk} + \hat{U}_{\vec{n},jk}^\dag\right].
    \label{eq:KogutSusskind}
\end{equation}

\section{(3+1)D topological terms in the Hamiltonian formulation}
\label{sec:top}

In this section, we derive a lattice definition of the (3+1)D topological $\theta$-term in the Hamiltonian formulation, using the transfer matrix method~\cite{creutz1977gauge}. This novel derivation complements the well-know known lattice definition of the (3+1)D $\theta$-term in the Lagrangian formulation (see Ref.~\cite{Alexandrou:2017hqw} for a review). 

In terms of the field strength tensor, the continuum topological $\theta$-term in the Lagrangian formulation reads
\begin{align}
\theta Q(x)&= \frac{\theta g^2}{16\pi^2}\sum_{\mu, \nu }F_{\mu\nu}(x)\tilde{F}_{\mu\nu}(x) \nonumber \\
         &= \frac{\theta g^2}{32\pi^2}\sum_{\mu,\nu,\rho,\sigma}\varepsilon_{\mu\nu\rho\sigma}F_{\mu\nu}(x)F_{\rho\sigma}(x) \nonumber \\
         &= \frac{\theta g^2}{8\pi^2}\sum_{i,j,k}\varepsilon_{ijk} F_{0i}(x) F_{jk}(x) \nonumber \\
         &= \frac{\theta g^2}{8\pi^2}\sum_{i,j,k,b}\varepsilon_{ijk} \text{Tr}\left[F_{0i}^b(x) F_{jk}^b(x)\right],
\label{eq:theta_term}
\end{align}
where $Q(x)$ is the topological charge,  $\varepsilon_{\mu\nu\rho\sigma}$ is the 4D Levi-Civita symbol, $F_{\mu\nu}(x) = \sum_b F_{\mu\nu}^b(x) \lambda^b$ is the field strength tensor, $\tilde{F}_{\mu\nu}(x)=\frac{1}{2}\sum_{\rho,\sigma} \varepsilon_{\mu\nu\rho\sigma}F_{\rho\sigma}(x)$ is the Hodge dual of $F_{\mu\nu}(x)$, and $\theta$ is an angular variable that can be shifted by $\theta\to\theta+n2\pi$, $n\in \mathds{Z}$, keeping the Lagrangian invariant. In the third equality, we used the fact that both the Levi-Civita symbol and the field strength tensor gain minus signs when exchanging two of their indices, which cancel out. 

In order to derive the generic structure of the $\theta$-term on the lattice in the Hamiltonian formulation, one can start with Peskin's original lattice definition of the topological charge in the Lagrangian formulation~\cite{Peskin1978,DiVecchia:1981aev},
\begin{align}
Q_{\vec{n}} &= -\frac{1}{32\pi^2} \sum_{\mu,\nu,\rho,\sigma}\varepsilon_{\mu\nu\rho\sigma}\,\textrm{Tr}\, \left[ U_{\vec{n},\mu\nu} U_{\vec{n},\rho\sigma}\right],
\label{eq:LTopCharge1}
\end{align}
and perform an expansion of the first plaquette variable,
\begin{align}
 Q_{\vec{n}}= - &\frac{1}{32\pi^2}\,\sum_{\mu,\nu,\rho,\sigma,b} \varepsilon_{\mu\nu\rho\sigma}\nonumber \\
 &\textrm{Tr}\, \left[\left(1+iga_\mu a_\nu F^b_{\vec{n},\mu\nu}\lambda^b+\ldots\right)U_{\vec{n},\rho\sigma}\right]\nonumber\\
= - &\frac{1}{16\pi^2}\sum_{i,j,k,b}\varepsilon_{ijk}\, \nonumber\\
&\textrm{Tr}\, \left[\left(1+iga a_0\left(F^b_{\vec{n},0i}-F^b_{\vec{n},i0}\right)\lambda^b+\ldots\right)U_{\vec{n},jk}\right]\nonumber\\
= - &\frac{1}{8\pi^2}\sum_{i,j,k,b}\varepsilon_{ijk}\,\textrm{Tr}\, \left[\left(1+iga a_0 E_{\vec{n},i}^b\lambda^b+\ldots\right)U_{\vec{n},jk}\right]\nonumber\\
= - &\frac{iga a_0}{8\pi^2}\sum_{i\perp j \perp k, b}\textrm{Tr}\,\left[E_{\vec{n},i}^b\lambda^b\left(U_{\vec{n},jk}-U^\dagger_{\vec{n},jk}\right)+\ldots\right],
\label{eq:LTopCharge2}
\end{align}
where $a_\mu=a_0$ for $\mu=0$, and $a_\mu=a$ otherwise. At first sight, it seems as if this naive expansion does not yield the correct prefactor for the $\theta$-term in the Hamiltonian formulation. Similarly, it seems as if a naive expansion of Wilson's lattice action in Eq.~\eqref{eq:Wilson} does not yield the correct prefactors for the electric and magnetic terms of the Kogut-Susskind Hamiltonian in Eq.~\eqref{eq:KogutSusskind}. From the transfer matrix derivation in Sec.~\ref{sec:lattice} we know that the magnetic term in the Hamiltonian formulation has the same $1/g^2$ prefactor as in the Lagrangian formulation, but the electric term acquires a $g^2$ prefactor through the Gaussian integration in Eq.~\eqref{eq:Gaussian}. This apparent deviation can be explained by the fact that the electric fields in the Hamiltonian and Lagrangian formulations are defined differently.\footnote{We thank Artur Avkhadiev for pointing this out.} In particular, these two electric fields are related via $\hat{E}^{b}_{\vec{n},i} = (a^2/g){E}^{b}_{\vec{n},i}$ \cite{PhysRevD.102.094515}. In order to demonstrate that this is indeed the case [see Eq.~\eqref{eq:top1}], we will derive the topological $\theta$-term using the transfer matrix method in the following.

The Euclidean lattice action, including the topological charge in Eq.~\eqref{eq:LTopCharge1} with a vacuum angle $\theta$, is~\cite{bhanot1984lattice}
\begin{equation}
    S=S_W+i\theta \sum_{\vec{n}} Q_{\vec{n}},
\end{equation}
where the topological charge picks up a factor of $i$ when going from Minkowski to Euclidean spacetime~\cite{bilal2008lectures}.

Just as in the second line of Eq.~\eqref{eq:LTopCharge2}, we isolate the temporal components of $Q_{\vec{n}}$ in the action,
\begin{align}
    \quad &\text{Tr}\left[ U_{\vec{n},0i} U_{\vec{n},jk}- U_{\vec{n},i0} U_{\vec{n},jk}\right] \nonumber \\
    =\, &\text{Tr}\left[ U_{\vec{n},0i} U_{\vec{n},jk}- U_{\vec{n},0i}^\dag U_{\vec{n},jk}\right] \nonumber \\
    =\, &\text{Tr}\left[(U_{\vec{n},0i} - U_{\vec{n},0i}^\dag)U_{\vec{n},jk}\right],
\end{align}
such that we can write the action as
\begin{align}
    S &= S_W - \frac{i\theta}{16\pi^2}\sum_{\vec{n},i,j,k}\varepsilon_{ijk}\text{Tr}\left[\left(U_{\vec{n},0i} - U_{\vec{n},0i}^\dag\right)U_{\vec{n},jk}\right].
    \label{eq:Smod}
\end{align}
Working in the temporal gauge, we write the transfer matrix elements, which satisfy Eq.~\eqref{eq:partition2}, as
\begin{align}
\begin{split}
    \bra{U'} \hat{T} \ket{U}& = e^{\frac{a}{2g^2 a_0} \sum_{\vec{n},i} \text{Tr}\left[U_{\vec{n},i}'U_{\vec{n},i}^\dag + U_{\vec{n},i}^{\dag'} U_{\vec{n},i}\right]} \\
    & \quad\times e^{\frac{i\theta}{16\pi^2}\sum_{\vec{n},i,j,k}\varepsilon_{ijk}\text{Tr}\left[\left(U_{\vec{n},i}'U_{\vec{n},i}^\dag - U_{\vec{n},i}^{\dag'}U_{\vec{n},i}\right)U_{\vec{n},jk}\right]}\\
    & \quad\times e^{\frac{a_0}{2g^2 a} \sum_{\vec{n},j,k}\text{Tr}\left[U_{\vec{n},jk} + U_{\vec{n},jk}^\dag\right] }.
    \end{split}
    \label{eq:step1}
\end{align}
In terms of the operators defined in Eqs.~\eqref{eq:usefulop1}--\eqref{eq:usefulop2}, we write the transfer matrix as
\begin{align}
    \hat{T} &= \prod_{\vec{n},i}\int_{g \in G}  dg_{\vec{n},i} \hat{R}_{\vec{n},i}(g_{\vec{n},i})e^{\frac{a}{2g^2 a_0}  \text{Tr}\left[g_{\vec{n},i} + g_{\vec{n},i}^\dag\right]} \nonumber \\
    & \quad \times e^{\frac{i\theta}{16\pi^2}\sum_{j,k}\varepsilon_{ijk}\text{Tr}\left[\left(g_{\vec{n},i} - g_{\vec{n},i}^\dagger\right)\hat{U}_{\vec{n},jk}\right]} \nonumber \\
    &\quad \times e^{\frac{a_0}{2g^2 a} \sum_{\vec{n},j,k}\text{Tr}\left[\hat{U}_{\vec{n},jk} + \hat{U}_{\vec{n},jk}^\dag\right]}\nonumber \\
    &= \prod_{\vec{n},i}\int \left(\prod_{b} dx_{\vec{n},i}^b\right) e^{i\sum_b x_{\vec{n},i}^b \hat{E}^{b}_{\vec{n},i}} \nonumber \\
    &\quad \times e^{\frac{a}{2g^2 a_0} \text{Tr}\left[2\cos(\sum_b x_{\vec{n},i}^b\lambda^b)\right]}\nonumber\\
    & \quad \times e^{\frac{i\theta}{16\pi^2}\sum_{j,k}\varepsilon_{ijk}\text{Tr}\left[2i\sin\left(\sum_b x_{\vec{n},i}^b \lambda^b\right)\hat{U}_{\vec{n},jk}\right]} \nonumber \\
    & \quad \times e^{\frac{a_0}{2g^2 a} \sum_{\vec{n},j,k}\text{Tr}\left[\hat{U}_{\vec{n},jk} + \hat{U}_{\vec{n},jk}^\dag\right] }.
\end{align}
In the continuum limit, as $a_0 \rightarrow 0$, the integral is dominated by the maximum of the cosine term. Expanding the cosine and sine terms around $x_{\vec{n},i}^b = 0$, we obtain
\begin{align}
    \hat{T}&\approx \prod_{\vec{n},i} \int \left(\prod_{b} dx_{\vec{n},i}^b\right) e^{i\sum_b x_{\vec{n},i}^b \hat{E}^{b}_{\vec{n},i}-\frac{a}{2g^2 a_0} \sum_b x_{\vec{n},i}^b x_{\vec{n},i}^b} \nonumber \\
    & \quad \times e^{\frac{-\theta}{8\pi^2}\sum_{j,k,b}\varepsilon_{ijk}x_{\vec{n},i}^b\text{Tr}\left[ \lambda^b\hat{U}_{\vec{n},jk}\right]} \nonumber \\
    & \quad \times e^{\frac{a_0}{2g^2 a} \sum_{\vec{n},j,k}\text{Tr}\left[\hat{U}_{\vec{n},jk} + \hat{U}_{\vec{n},jk}^\dag\right] }.
    \label{eq:step4}
\end{align}
This Gaussian integral evaluates to
\begin{equation}
    \hat{T} \propto e^{-a_0\set*{\hat{H}_{KS}-\frac{ig^2 \theta}{8\pi^2 a}\sum_{\vec{n},i,j,k,b}\varepsilon_{ijk}\text{Tr}\left[\hat{E}_{\vec{n},i}^b \lambda^b\hat{U}_{\vec{n},jk}\right] }}.
    \label{eq:TransferResult}
\end{equation}
We note that at first sight, it seems as if the Gaussian integration yields an $\Ol(\theta^2)$ term, which comes from squaring the exponent in the second line of Eq.~\eqref{eq:step4}. However, this term is proportional to $\sum_{\vec{n},i,j,k,l,m}\varepsilon_{ijk}\varepsilon_{ilm}\text{Tr}[\lambda^b\hat{U}_{\vec{n},jk}]\text{Tr}[\lambda^b\hat{U}_{\vec{n},lm}]$, where $i$ is an index of the integral variable. This term cancels exactly due to $\sum_{i} \varepsilon_{ijk}\varepsilon_{ilm} = \delta_{jl}\delta_{km} - \delta_{jm}\delta_{kl}$. Thus, we obtain the topological $\theta$-term in the Hamiltonian formulation,\footnote{While our numerical calculations were already running, we became aware of a recent arXiv paper~\cite{cohen2021quantum} that independently derived a result similar to Eq.~\eqref{eq:top1} in a different context. The authors extract elements from bilinear field strength tensor combinations to study transport coefficients. While our derivation is based on the standard transfer matrix technique and does not rely on introducing the $\theta$-term as a small perturbativon, their perturbative method can in principle also be used to derive the $\theta$-term. We added a comparison between our derivation and the one in Ref.~\cite{cohen2021quantum} to App.~\ref{sec:altderivation}.}
\begin{align}
    \theta\hat{Q}&=-\frac{ig^2 \theta}{8\pi^2 a}\sum_{\vec{n},i,j,k,b}\varepsilon_{ijk}\text{Tr}\left[\hat{E}_{\vec{n},i}^b \lambda^b\hat{U}_{\vec{n},jk}\right]
    \nonumber \\
    &= -\frac{ig^2 \theta}{4\pi^2 a}\sum_{\vec{n},b} \sum_{(i,j,k)\in \text{even}}\text{Tr}\left[\hat{E}_{\vec{n},i}^b \lambda^b\left(\hat{U}_{\vec{n},jk}-\hat{U}_{\vec{n},jk}^\dag\right)\right].
    \label{eq:top1}
\end{align}
In the last line, $(i,j,k)$ is summed over the set of even permutations, and we have used $\hat{U}_{\vec{n},jk} = \hat{U}_{\vec{n},kj}^\dag$.

To improve the definition in Eq.~\eqref{eq:top1}, for each site $\vec{n}$, we replace the outgoing electric field $\hat{E}_{\vec{n},i}^b$ with an average of the incoming and outgoing electric fields, which better approximates the field at each site, and obtain
\begin{multline}
    \theta \hat{Q} = -\frac{ig^2 \theta}{8\pi^2 a}\sum_{\vec{n},b} \sum_{(i,j,k)\in \text{even}} \\
    \text{Tr}\left[\left(\hat{E}_{\vec{n}-\hat{i},i}^b+\hat{E}_{\vec{n},i}^b\right) \lambda^b\left(\hat{U}_{\vec{n},jk}-\hat{U}_{\vec{n},jk}^\dag\right)\right].
    \label{eq:TopTermSymm}
\end{multline}
Note that the $\theta$-term is not invariant under the CP transformation $\hat{U}_{\vec{n},i}\rightarrow \hat{U}_{\vec{n},i}^\dag$, since the topological charge changes its sign. This is due to the totally antisymmetric $\varepsilon_{ijk}$ symbol in Eq.~\eqref{eq:top1}, which changes its sign when reversing two indices, corresponding to a parity transformation. This CP violation manifests in the pseudovector nature of the magnetic field, $B_i = -\frac{1}{2} \varepsilon_{ijk} F_{jk}$, which appears in the $\theta$-term, $\theta Q\propto \varepsilon_{\mu\nu\rho\sigma}F_{\mu\nu}F_{\rho\sigma}\propto\mathbf{E}\cdot \mathbf{B}$. As explained in the introduction, the CP violating nature of the $\theta$-term is the origin of the strong CP problem, which is the problem that QCD does not seem to distinguish matter from antimatter~\cite{Hook:2018dlk}.

Since continuous gauge groups lead to infinite-dimensional Hilbert spaces, in order to simulate the Hamiltonian on a finite-size classical or quantum computer, it is necessary to truncate these to render the problem finite-dimensional. For any finite or compact Lie group, this can be accomplished by expanding the electric field and link operators in the group representation basis, and truncating the irreducible representation labels. This method is detailed in Ref.~\cite{zohar2015formulation}. In this work, we choose to focus on the simplest non-trivial truncation of the U(1) gauge group.

\section{Model and methods}
\label{sec:model}

\begin{figure*}[hbt!]
    \includegraphics[width=\textwidth]{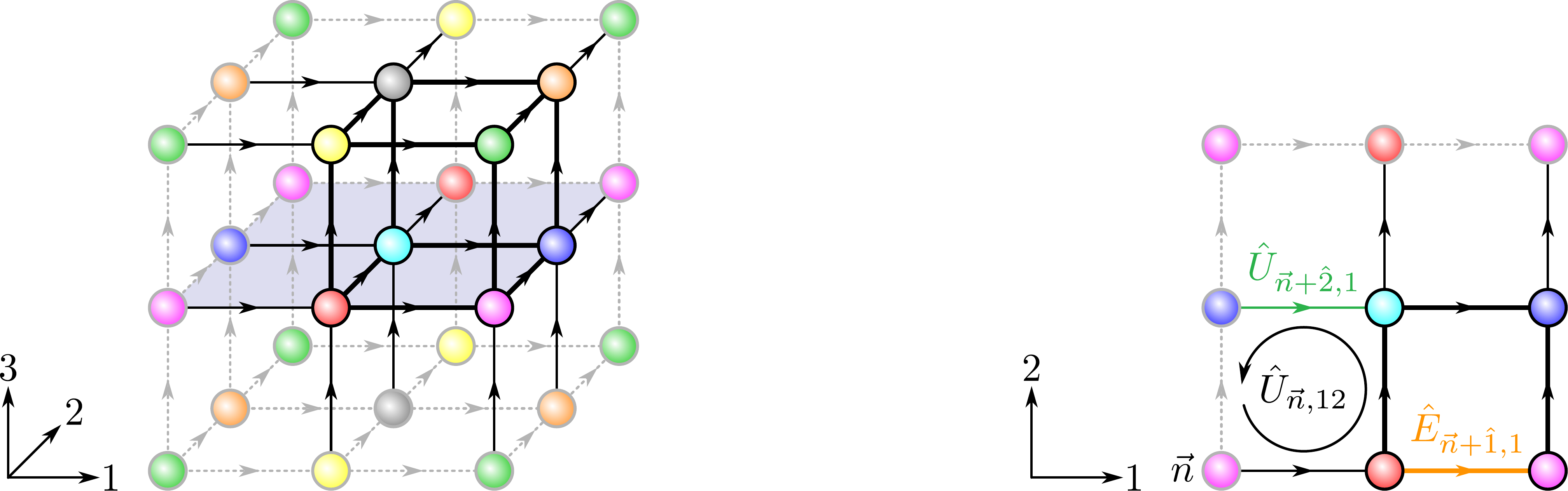}
    \caption{Left: Sketch of the 3D cube with periodic boundary conditions. The upper right corner shows the cube with bold black lines, where the colored circles with black outlines correspond to the 8 different vertices at the corners. To illustrate the periodic boundary conditions, the original 8 vertices are mirrored in every direction with mirrored vertices indicated by the same color as the original ones but with grey outlines. The 24 links are indicated as solid black lines with the arrows indicating the orientation of the links. The dashed grey lines correspond to mirrored links due to the periodic boundary conditions.\\ Right: Illustration of a cut through the middle layer of the cube along the $1$-$2$ plane, which is highlighted in light blue in the left panel. For illustration purposes, we show a link operator (green), an electric field operator (orange), and a plaquette operator (black) corresponding to the product of the link operators around the plaquette as indicated by the circular arrow.}
    \label{fig:periodic_cube}
\end{figure*}

As a particular example for the generic expression derived in Eq.~\eqref{eq:TopTermSymm}, we now numerically investigate a (3+1)D U(1) lattice gauge theory. We use a single cube with periodic boundary conditions (see Fig.~\ref{fig:periodic_cube}) and explore the theory in the Hamiltonian formulation at non-vanishing $\theta$ using exact diagonalization. In our computations, we set the lattice spacing $a=1$ and consider the Hamiltonian
\begin{align}
    \hat{H} &= \hat{H}_E + \hat{H}_B + \tilde{\theta} \hat{Q}, \label{eq:H_sim} \\
   \hat{H}_E &= \frac{1}{2\beta}\sum_{\vec{n}}\sum_{j=1}^{3} \hat{E}_{\vec{n},j}^2, \label{eq:HE} \\
    \hat{H}_B &= -\frac{\beta}{2} \sum_{\vec{n}}\sum_{j,k=1;k>j}^{3} (\hat{U}_{\vec{n},jk} + \hat{U}_{\vec{n},jk}^\dag )\label{eq:HB},\\
    \tilde{\theta} \hat{Q} &= -i \frac{\tilde{\theta}}{\beta} \sum_{\vec{n}}\sum_{(i,j,k)\in \text{even}} (\hat{E}_{\vec{n}-\hat{i},i}+\hat{E}_{\vec{n},i})(\hat{U}_{\vec{n},jk} - \hat{U}_{\vec{n},jk}^\dag), \label{eq:U1_Q}
\end{align}
where $\tilde{\theta} = \theta/8\pi^2$, $\beta = 1/g^2$, $(i,j,k)$ is summed over the set of even permutations, and the link operators satisfy the commutation relation
\begin{equation}
    [\hat{E}_{\vec{n},j}, \hat{U}_{\vec{n}',j'}] = \delta_{\vec{n},\vec{n}'}\delta_{j,j'} \hat{U}_{\vec{n},j}.
    \label{eq:U1_comm}
\end{equation}
The eigenstates of the electric field operators
\begin{equation}
    \hat{E}_{\vec{n},j}\ket{{E}_{\vec{n},j}}= {E}_{\vec{n},j}\ket{{E}_{\vec{n},j}}, \: {E}_{\vec{n},j} \in \mathbbm{Z}
\end{equation}
form a basis for the Hilbert space of the gauge fields. In this basis, the gauge field operators can be represented as
\begin{gather}
    \hat{E}_{\vec{n},j} = \sum_{{E}_{\vec{n},j} \in \mathbbm{Z}}{E}_{\vec{n},j}\ketbra{{E}_{\vec{n},j}}{{E}_{\vec{n},j}}, \\
    \hat{U}_{\vec{n},j} = \sum_{{E}_{\vec{n},j} \in \mathbbm{Z}}\ketbra{{E}_{\vec{n},j}-1}{{E}_{\vec{n},j}}.
\end{gather}
It can be checked straightforwardly that these operators satisfy the commutation relation in Eq.~\eqref{eq:U1_comm}. In order to represent the infinite-dimensional gauge-field operators on a finite-size computer, their Hilbert space must be truncated at a cutoff, $s$. Thus, the gauge field operators become
\begin{gather}
    \hat{E}_{\vec{n},j} = \sum_{{E}_{\vec{n},j}=-s}^{s}{E}_{\vec{n},j}\ketbra{{E}_{\vec{n},j}}{{E}_{\vec{n},j}}, \\
    \hat{U}_{\vec{n},j} = \sum_{{E}_{\vec{n},j}=-s+1}^{s}\ketbra{{E}_{\vec{n},j}-1}{{E}_{\vec{n},j}}.
\end{gather}
Throughout this work, we choose the simplest nontrivial symmetric truncation corresponding to $s=1$.

Furthermore, the Hamiltonian is gauge invariant because it commutes with the Gauss' law operators
\begin{equation}
    \hat{G}_{\vec{n}} = \sum_{i = 1}^{3}(\hat{E}_{\vec{n},i}-\hat{E}_{\vec{n}-\hat{i},i}), \: \forall \vec{n}.
\end{equation}
The gauge-invariant physical states $\ket{\Psi}$ are constrained by the Gauss' law operators via the relation
\begin{equation}
    \hat{G}_{\vec{n}}\ket{\Psi} = 0, \: \forall \vec{n}.
    \label{eq:gauss}
\end{equation}
However, the Hamiltonian acts on a Hilbert space that contains many unphysical states, which violate the gauge-invariance condition in Eq.~\eqref{eq:gauss}. Therefore, the spectrum of the Hamiltonian, without enforcing Gauss' law, will be contaminated by unphysical states. A possible way to suppress the unphysical states is to diagonalize $\hat{H} + r \sum_{\vec{n}} \hat{G}_{\vec{n}}^2$ with $r \gg 1$, where the squared Gauss' law operators are included as a penalty term~\cite{hauke2013,kuhn2014}. Since the physical states lie in the kernel of the Gauss' law operators, they will be unaffected by the penalty. We choose to use an alternative and more resource-efficient way to incorporate Gauss' law into our Hamiltonian, following Ref.~\cite{haase2021resource}. In particular, we treat Eq.~\eqref{eq:gauss} as a set of constraints on the electric operators
\begin{equation}
    \sum_{i = 1}^{3}(\hat{E}_{\vec{n},i}-\hat{E}_{\vec{n}-\hat{i},i}) = 0, \: \forall \vec{n},
\end{equation}
and solve this as a system of equations over the electric operators. Since the sum of all the Gauss' law constraints evaluates to zero, $\sum_{\vec{n}=1}^{N} \hat{G}_{\vec{n}} = 0$, there are only $N-1$ independent constraints on a lattice with $N$ sites.
Hence, we can eliminate $N-1$ arbitrary electric field operators by expressing them in terms of the remaining ones~\cite{haase2021resource}. Since the eliminated electric field operators no longer contribute directly to the dynamics, their corresponding link operators become identities. In one dimension with open boundary conditions, this method allows one to completely eliminate the gauge fields, leaving only matter degrees of freedom~\cite{hamer1997,Banuls2013,Banuls2016a}. This method can be applied to higher dimensions, as discussed in Ref.~\cite{PhysRevD.102.014506,PhysRevD.102.094515,PhysRevD.99.074502}, and has recently been demonstrated on a (2+1)D lattice gauge theory~\cite{haase2021resource}. For a 3D cubic lattice with periodic boundary conditions, where $N=L^3$, $L^3-1$ out of $3L^3$ link degrees of freedom are eliminated, and thus, the Hamiltonian can be expressed solely in terms of the gauge field operators acting on the remaining $2L^3-1$ links. Here, $L$ denotes the number of sites along each direction. Compared to the penalty method, where the dimensions of the Hamiltonian remain unchanged, this method reduces the number of basis states from $(2s+1)^{3L^3}$ to $(2s+1)^{2L^3-1}$.

\section{Results}
\label{sec:results}

\begin{figure*}[htp!]
\begin{tabular}{cc}
  \includegraphics[width=0.48\linewidth]{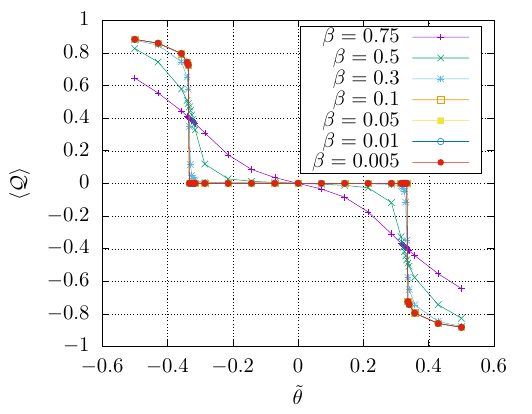} & \includegraphics[width=0.48\linewidth]{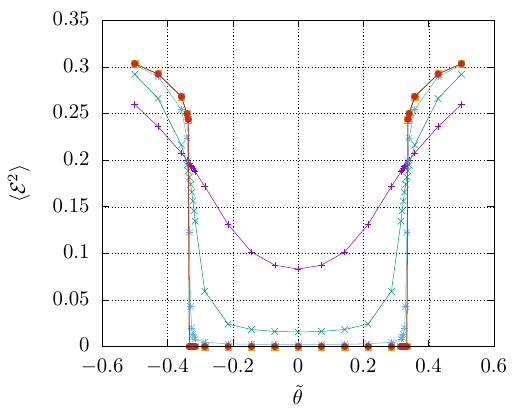}    \\
 (a) & (b) \\[6pt]
  \includegraphics[width=0.48\linewidth]{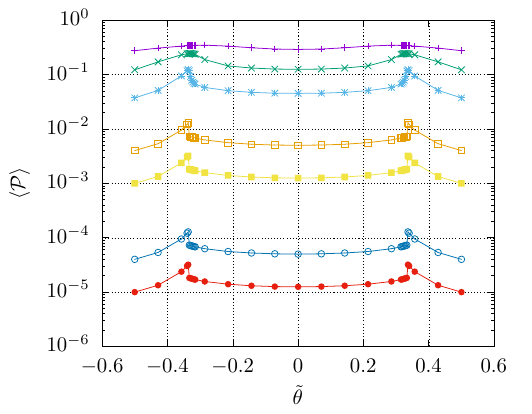}& \includegraphics[width=0.48\linewidth]{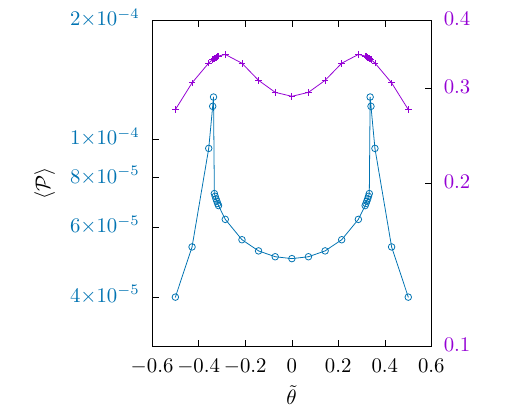}  \\ 
 (c) & (d) \\[6pt]
\end{tabular}
\caption{(a) Bare topological charge density, (b) bare electric energy density, and (c) plaquette expectation value as a function of $\tilde{\theta}$ for $\beta \leq 0.75$. (d) The plaquette expectation values for $\beta = 0.01$ (left $y$-axis) and $0.75$ (right $y$-axis) are shown in greater detail to highlight the change in the behavior as $\beta$ increases. Note that in panels (a) and (b) the lines for $\beta=0.1,0.05,0.01$  are covered by the red line.}
\label{fig:observables}
\end{figure*}

In order to investigate the zero-temperature phase structure of the U($1$) gauge theory in the presence of a topological term on a periodic cube, we compute the low-lying spectrum of the Hamiltonian in Eq.~\eqref{eq:H_sim} using exact diagonalization. In our numerical analysis, we focus on the $\theta$-dependence of the energy spectrum and the expectation value of various observables in the ground state $\ket{\Psi_0}$ of the Hamiltonian. In particular, we study the plaquette expectation value
\begin{equation}
    \expval{\mathcal{P}} = -\frac{1}{V\beta} \langle \Psi_0 |\hat{H}_B| \Psi_0\rangle,
    \label{eq:plaq}
\end{equation}
where $V$ is the number of plaquettes on the lattice, the bare topological charge density
\begin{equation}
    \expval{\mathcal{Q}} = -\frac{\beta}{V} \langle \Psi_0 |\hat{Q}| \Psi_0\rangle,
\end{equation}
the bare electric energy density
\begin{equation}
    \expval{\mathcal{E}^2} = \frac{\beta}{V} \langle \Psi_0 |\hat{H}_E| \Psi_0\rangle,
\end{equation}
and the electric field expectation value
\begin{equation}
 \expval{\mathcal{E}} = \langle \Psi_0 |\sum_{\vec{n},j} \hat{E}_{\vec{n},j}| \Psi_0\rangle .  
\end{equation}
We explore a wide range of couplings corresponding to $\beta \in [0.005,0.75]$, and for each value of the coupling a range of values $\tilde{\theta} \in [-0.5,0.5]$. 
Fig.~\ref{fig:observables} shows our results for the topological charge density, the electric energy density, and the plaquette expectation value. Focusing on the regime $\beta \leq 0.3$, we see that the topological charge density as well as the electric energy density show sharp discontinuities at $|\tilde{\theta}| \approx 0.333$, as shown in Figs.~\ref{fig:observables}(a) and \ref{fig:observables}(b). Similarly, the plaquette expectation value has distinct spikes at these points, as Figs.~\ref{fig:observables}(c) and \ref{fig:observables}(d) reveal. The sharp discontinuities in the electric energy density and the topological charge density, together with the spike in the plaquette expectation value indicate that a phase transition is occurring at $|\tilde{\theta}| \approx 0.333$. 

Going to smaller values of the coupling, or equivalently increasing $\beta$ beyond $0.3$, we can clearly see that the distinct features in all three observables eventually vanish and the curves become smooth. In particular, the spikes in the plaquette expectation value disappear, as the comparison in Fig.~\ref{fig:observables}(d) shows. Hence, our data does not show any signs of a phase transition in this regime.

Comparing our numerical results to the analytical findings of Refs.~\cite{cardy1982duality,honda2020topological}, we find qualitative agreement. These references predict periodic behavior in $\theta$ with a period of $2\pi$, and a phase transition at $\theta=\pi$ for large values of the coupling, or equivalently small $\beta$. This transition should eventually vanish as one approaches the Coulomb phase at small values of the coupling, corresponding to large $\beta$. The transition at $\theta=\pi$ is, to the best of our knowledge, the only transition that occurs at constant $\theta$ in a compact U(1) gauge theory. Hence, despite a significant shift towards larger values of $\theta$, it seems likely that the transition we observe corresponds to the theoretically predicted one at $\theta=\pi$. Moreover, while our  data is not perfectly periodic at large values of $\theta$, we see the following symmetries in the observables, i.e., $\expval{\mathcal{P}(\theta)} = \expval{\mathcal{P}(-\theta)}$, $\expval{\mathcal{E}^2(\theta)} = \expval{\mathcal{E}^2(-\theta)}$, and $\expval{\mathcal{Q}(\theta)} = -\expval{\mathcal{Q}(-\theta)}$. The oddness of the topological charge density and the evenness of the plaquette expectation and the electric energy density can be directly inferred from Eq.~\eqref{eq:U1_Q}. Indeed, the CP transformation $U_{\vec{n},i}\rightarrow U_{\vec{n},i}^\dag$ only changes the sign of the topological charge but leaves the plaquette expectation and the electric energy density unchanged. This leads to a linear behavior of $\expval{\mathcal{Q}(\theta)}$ near $\theta = 0$, as observed in Fig.~\ref{fig:observables}(a). Note that a similar behavior has also been observed in lower-dimensional lattice studies of topological terms, such  as Refs.~\cite{gattringer2015,Funcke:2019zna}.

The shift in the transition point and the lack of perfect periodicity in our data are likely to be caused by the truncation of the electric field and by the small volume we study. To probe how severely the electric field truncation affects our results, we can investigate the expectation value of the electric field $\expval{\mathcal{E}}$. In particular, for $\beta \leq 0.75$, we expect the ground state to have a significant overlap with the electric vacuum, and thus, yield $\expval{\mathcal{E}} \approx 0$. For all values of $\theta$ and $\beta$ explored in our numerical calculations, we indeed observe that $|\expval{\mathcal{E}}| < 7 \times 10^{-13}$. This suggests, that we are in a regime in which the electric field values are small, and that truncation effects play an insignificant role in our computations. 

In contrast, we expect finite-volume effects to have a larger impact. In particular, it has been observed in previous studies of the (1+1)D Abelian-Higgs model with a $\theta$-term that for very small volumes the model is only approximately periodic in $\theta$ with a period larger than $2\pi$~\cite{gattringer2015,higgs1d}. The perfect $2\pi$-periodicity is only restored on large lattices~\cite{gattringer2015,Funcke:2019zna,higgs1d}. Furthermore, Ref.~\cite{higgs1d} found that the predicted phase transition at $\theta=\pi$ is significantly shifted towards larger values of $\theta$ for small volumes. Once again, these shifts eventually disappear for larger volumes. For the (3+1)D model we study, we expect even stronger finite-volume effects to take place, since the connectivity of each lattice site is higher. This could explain why we do not observe a transition at $\theta=\pi$ but rather at $\tilde{\theta}\approx 0.333$, which is roughly a factor of 8 larger.
\begin{figure}[htp!]
    \includegraphics[width=\linewidth]{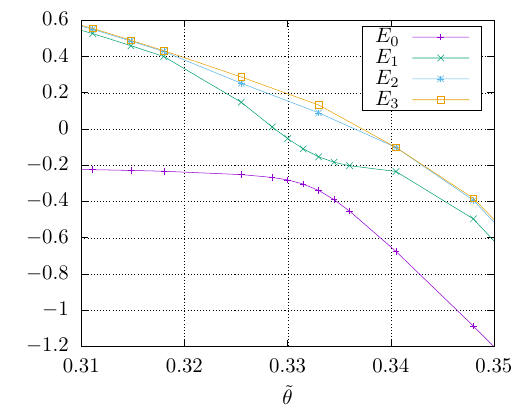}
    \caption{Low-lying spectrum as a function of $\tilde{\theta}$ for $\beta = 0.3$. The ground state and the first excited state show an avoided level-crossing at $\tilde{\theta} \approx 0.333$.}
    \label{fig:crossing}
\end{figure}

The nature of the phase transition at $\theta=\pi$ is unknown from analytical predictions~\cite{Cardy:1981qy,cardy1982duality}. To get further insights into the transition that we observe, we can study the low-lying energy spectrum of the Hamiltonian. We focus on $\beta=0.3$, at which we still see clear signatures of a phase transition, but which is already close to the point where the abrupt changes in the observables begin to smoothen out. Fig.~\ref{fig:crossing} shows the results for the first four energy levels of the Hamiltonian close to the observed transition point $\tilde{\theta} \approx 0.333$. As the figure reveals, there is an avoided level-crossing between the ground state and the first excited state at the transition point. In particular, we do not find a level crossing as one would expect for a first-order quantum phase transition. The low-lying spectrum rather hints towards a second or higher-order phase transition causing the features we see in Fig.~\ref{fig:observables} in the topological charge density, the electric energy density, and the plaquette expectation value for small $\beta$. Furthermore, the theory at $\tilde{\theta}\approx 0.333$ would correspond to one with $\theta=0$ and a negative fermionic mass once the U($1$) gauge theory is coupled to a fermion (see App.\ \ref{sec:(3+1)Danomaly} for more details on this correspondence). Future investigations of this claim can be accomplished using quantum simulations and tensor network simulations.

\section{Conclusion and Discussion}
\label{sec:conclusion}

In this paper, we derived the topological $\theta$-term for (3+1)D Abelian and non-Abelian lattice gauge theories in the Hamiltonian formulation, using the transfer matrix method. We then applied our derivation to the (3+1)D pure compact U(1) lattice gauge theory and explored the topological zero-temperature phase structure using exact diagonalization for a single periodic cube. In our numerical calculations, we obtained evidence for a phase transition at constant values of $\theta$ and for large values of the coupling $g$. The transition appears in form of an avoided level-crossing in the spectrum, discontinuities in electric energy density, and topological charge density, and a spike in the plaquette expectation value. Our data suggests that the observed transition corresponds to the analytically predicted one at $\theta=\pi$ for large values of $g$. The low-lying energy spectrum indicates that this transition is not of first order. Decreasing the coupling and moving towards the Coulomb phase, we observe that the transition eventually vanishes, in line with the theoretical predictions of Refs.~\cite{cardy1982duality,honda2020topological}.  

Our work enables future investigations of topological effects in lattice gauge theories via quantum and classical algorithms, which are formulated in terms of the Hamiltonian. On the classical front, our numerical results can be cross-checked and extended to larger lattices using tensor network methods developed for (3+1)D lattice gauge theories in Ref.~\cite{Magnifico:2020bqt}. Furthermore, it will be interesting to explore the $\theta$-dependent phase structure near the de-confinement transition at $\beta \approx 1$ \cite{bhanot1980phase,jersak1983u}. For simulations of small lattices, one can include the extended action described in Ref.~\cite{bhanot1982compact} to enhance the de-confinement transition and to counteract finite-size effects. The extended Hamiltonian corresponding to the extended action is derived in App.~\ref{sec:extended}. Furthermore, using the magnetic basis devised in Ref.~\cite{haase2021resource}, one can further explore the phase diagram in the Coulomb phase, where $\beta > 1$, with a modest amount of computational resources and minimal truncation effects. On the quantum front, our derivations provide a necessary starting point for designing quantum algorithms to simulate models with a topological $\theta$-term. Furthermore, our results demonstrate that even a minimal (3+1)D lattice gauge theory formulated on a single cube can already exhibit a non-trivial topological phase structure. This is a valuable insight for the development of quantum simulators, as it implies a low overhead to realize interesting physical phenomena. Finally, our work can be extended by constructing a finite-dimensional representation of the $\theta$-term in non-Abelian theories for quantum simulations, using the method described in Ref.~\cite{zohar2015formulation}.

\section*{Acknowledgements}

L.F.\ thanks Artur Avkhadiev, Mike Creutz and Dima Kharzeev for discussions.  C.A.M.\ acknowledges the Alfred P. Sloan foundation for a Sloan Research Fellowship.
S.K.\ acknowledges financial support from the Cyprus Research and Innovation Foundation under project “Future-proofing Scientific Applications for the Supercomputers of Tomorrow (FAST)”, contract no.\ COMPLEMENTARY/0916/0048. J.F.H.\ acknowledges the Alexander von Humboldt Foundation in the form of a Feodor Lynen Fellowship.
This work is supported in part by Transformative Quantum Technologies Program (CFREF), NSERC, New frontiers in Research Fund, Compute Canada, and European Union’s Horizon 2020 research and innovation programme under the Grant Agreement No.\ 731473 (FWF QuantERA via QTFLAG I03769). Research at Perimeter Institute is supported in part by the Government of Canada through the Department of Innovation, Science and Industry Canada and by the Province of Ontario through the Ministry of Colleges and Universities. 
\appendix

\section{Alternative transfer matrix derivation of the topological \texorpdfstring{\boldmath{$\theta$}-}{}term}\label{sec:altderivation}

In this appendix, we show that using the analytical method recently proposed in Ref.~\cite{cohen2021quantum}, one can derive the same expression for the topological $\theta$-term in the Hamiltonian formulation as we derived in Eq.~\eqref{eq:top1}. The authors of Ref.~\cite{cohen2021quantum} derived the Hamiltonian plaquette formulation of $\sum_b F_{\vec{n},0j}^b F_{\vec{n},jk}^b$ in the context of studying transport coefficients, and we became aware of their results only recently, while running our numerical calculations.

In Ref.~\cite{cohen2021quantum}, the authors introduce different bilinear combinations of the field strength tensor as a perturbation to Wilson's lattice action $S$, from which they derive the operators that corresponds to the variables. While we adopt Euclidean spacetime in the main text, Minkowski spacetime is adopted in the following, as in Ref.~\cite{cohen2021quantum}. Briefly, if the perturbed action is $S + \epsilon O$, where $O$ is the perturbation, the perturbed transfer matrix elements are
\begin{equation}
    \bra{U'} e^{-ia_0 \hat{H}'} \ket{U} = e^{i(S+\epsilon O)},
\end{equation}
where $\hat{H}'$ is the perturbed Hamiltonian. Differentiating the right hand side with respect to $\epsilon$ yields the operator corresponding to $O$, up to a factor of $i$. Thus, the perturbed Hamiltonian is
\begin{equation}
    \hat{H}' = \hat{H}_{KS} - \frac{\epsilon}{a_0} \hat{O}.
    \label{perturbedO}
\end{equation}
As such, one can add the topological term to Wilson's lattice action as a perturbation
\begin{align}
    S' &= S_W + \epsilon a_0 a^3  \sum_{\vec{n},i,j,k,b} \text{Tr}\left[F_{\vec{n},0i}^b F_{\vec{n},jk}^b\right] \nonumber \\
    &=S_W - \frac{\epsilon}{4}\sum_{\vec{n},i,j,k}\text{Tr}\left[\left(U_{\vec{n},0i} - U_{\vec{n},0i}^\dag\right)\left(U_{\vec{n},jk} - U_{\vec{n},jk}^\dag\right)\right] \nonumber \\
    &\quad + \Ol(a),
    \label{eq:perturbedS}
\end{align}
using the relation
\begin{equation}
    \frac{-i}{2g a_\mu a_\nu} \left(U_{\vec{n},\mu\nu} - U_{\vec{n},\mu\nu}^\dag\right) = \sum_b F^b_{\vec{n}, \mu\nu}\lambda^b + \Ol(a),
    \label{eq:expansion}
\end{equation}
where $a_\mu = a$ if $\mu \neq 0$. One can see that the action in Eq.~\eqref{eq:perturbedS} is equivalent to the action in Eq.~\eqref{eq:Smod} that we used for our derivation, up to $\Ol(a)$ discretization error, a factor of $\epsilon$, and the Levi-Civita symbol. We note that our method does not require the topological term to be introduced as a perturbation.

The transfer matrix can be obtained following the steps in Eqs.~\eqref{eq:step1}--\eqref{eq:step4} and evaluates to
\begin{equation}
    \hat{T} \propto e^{-ia_0\set*{\hat{H}_{KS}-\frac{i\epsilon}{2a}\sum_{\vec{n},i,j,k,b}\text{Tr}\left[\hat{E}_{\vec{n},i}^b \lambda^b\left(\hat{U}_{\vec{n},jk}-\hat{U}_{\vec{n},jk}^\dagger \right)\right] + \Ol\left(\epsilon^2\right)}}.
\end{equation}
Keeping terms up to $\Ol(\epsilon)$ and using Eq.~(\ref{perturbedO}), one can identify the operator that corresponds to the perturbation in Eq.~\eqref{eq:perturbedS} as
\begin{equation}
    \frac{i}{2a}\sum_{\vec{n},i,j,k,b}\text{Tr}\left[\hat{E}_{\vec{n},i}^b \lambda^b\left(\hat{U}_{\vec{n},jk}-\hat{U}_{\vec{n},jk}^\dagger \right)\right],
    \label{eq:top2}
\end{equation}
which is structurally similar to Eq.~\eqref{eq:top1}. Note that when applied to the $\theta$-term, the removal of the $\Ol(\epsilon^2)$ does not rely on the fact that the $\theta$-term is a perturbation, but is rather due to the totally antisymmetric nature of the Levi-Civita symbol in Eq.~\eqref{eq:top1}.

\section{Equivalence of negative fermionic mass and topological \texorpdfstring{\boldmath{$\theta$}-}{}term with \texorpdfstring{\boldmath{$\theta=\pi$}}{}}
\label{sec:(3+1)Danomaly}
In this appendix, we discuss the well-known equivalence of a gauge field theory with a negative fermionic mass term and the same theory with a positive mass term plus a topological $\theta$-term at $\theta =\pi$. This equivalence is given because the $\theta$-parameter can be rotated from the $\theta$-term into the fermionic mass, $m\to m\exp(i\theta)$, which changes the sign of the mass for $\theta =\pi$, $m\to m\exp(i\pi)=-m$~\cite{Adler1969,Bell1969,Fujikawa1979,Dunne1989}. This equivalence holds true for any gauge theory with a non-trivial topological vacuum structure, such as the compact U(1) gauge theory or (compact or non-compact) non-Abelian gauge theories, including SU(2) and SU(3). Thus, in all of these theories, we expect a rich phase structure, in particular a phase transition at some critical coupling and negative mass corresponding to $\theta =\pi$.

To demonstrate the equivalence of a negative mass and a $\theta$-term at $\theta =\pi$, we need to consider the quantum anomaly equation of the continuum theory in the Hamiltonian formalism. For a massless fermion field $\hat{\psi}$, we can perform an chiral rotation of $\hat{\psi}$ without changing the classical Hamiltonian,
\begin{align}
\hat{\psi}&\rightarrow  e^{i\alpha\gamma^5}\hat{\psi}
,\label{eq:rot}
\end{align}
where $\alpha \in [0, 2 \pi ] $ is an angular variable. Thus, this rotation in Eq.~\eqref{eq:rot} is a symmetry of the classical Hamiltonian. However, this symmetry only holds true on the classical level and is violated on the quantum level.

If the rotation in Eq.~\eqref{eq:rot} were a true symmetry of the quantum Hamiltonian, the divergence of the fermionic current corresponding to this symmetry would vanish,
\begin{equation}
\sum_{\mu} \partial_\mu \hat{j}_5^\mu=0,
\end{equation}
where $\hat{j}_5^\mu=\hat{\bar{\psi}}\gamma^\mu\gamma^5\hat{\psi}$ is the chiral fermion current.
However, the presence of the quantum anomaly makes this term non-vanishing, which yields the non-trivial quantum anomaly equation for a massless fermion,
\begin{equation}
\sum_{\mu} \partial_\mu \hat{j}_5^\mu= \frac{g^2}{8\pi^2}\sum_{\mu,\nu} \hat{F}^{\mu\nu}\hat{\tilde{F}}_{\mu\nu}\label{eq:anomaly}
\end{equation}
where $g$ is the coupling, $\hat{{F}}^{\mu\nu}$ is the field strength tensor, and $\hat{\tilde{F}}^{\mu\nu}=\frac{1}{2}\sum_{\rho,\sigma} \varepsilon^{\mu\nu\rho\sigma}\hat{F}_{\rho\sigma}$ is its Hodge dual. This anomaly equation was first derived in the Lagrangian formalism, where instead of operators one has classical variables~\cite{Adler1969,Bell1969,Fujikawa1979}, and it is equivalent to the one in the Hamiltonian formalism~\cite{Dunne1989}. Note that in contrast to the main text, here we work in Minkowski spacetime. Thus, we use the usual convention of having upper and lower Lorentz indices, which are related by the metric tensor.

Now let us see what happens if we introduce a non-zero mass for the fermion $\hat{\psi}$. First, we observe that the rotation in Eq.~\eqref{eq:rot} shifts the fermionic mass term by
\begin{align}
\begin{split}
m\hat{\bar{\psi}}\hat{\psi}&\rightarrow m\hat{\bar{\psi}}e^{2i\alpha\gamma^5}\hat{\psi} \\
&\: = m\cos(2\alpha)\hat{\bar{\psi}}\hat{\psi} + i\,m\sin(2\alpha)\hat{\bar{\psi}}\gamma^5\hat{\psi}\\
&\: \approx m\hat{\bar{\psi}}\hat{\psi} + i\, 2\alpha m\hat{\bar{\psi}}\gamma_5\hat{\psi},
\end{split}
\label{eq:mass}
\end{align}
where the approximation holds true for $\alpha\ll 1$. 

Second, the fermionic mass perturbatively corrects the quantum anomaly equation in Eq.~\eqref{eq:anomaly} by
\begin{equation}
\sum_{\mu}\partial_\mu \hat{j}_5^\mu=\frac{g^2}{8\pi^2} \sum_{\mu,\nu}\hat{F}^{\mu\nu}\hat{\tilde{F}}_{\mu\nu}+i\,2m\hat{\bar{\psi}}\gamma_5\hat{\psi}.\label{eq:anomaly2}
\end{equation}
Using Eq.~\eqref{eq:anomaly2}, we can now demonstrate that the chiral rotation in Eq.~\eqref{eq:rot} by a small angle $\alpha\ll 1$ induces the following shift in the Hamiltonian:
\begin{align}
\begin{split}
\hat{H} &\rightarrow \hat{H} + \alpha\sum_{\mu} \partial_\mu \hat{j}_5^\mu\\
&\: =\hat{H}+ \frac{\alpha g^2}{8\pi^2} \sum_{\mu,\nu} \hat{F}^{\mu\nu}\hat{\tilde{F}}_{\mu\nu} + i\,2\alpha m\hat{\bar{\psi}}\gamma_5 \hat{\psi}.
\end{split}
\label{eq:H}
\end{align}
As expected, the last term in Eq.~\eqref{eq:H} coincides with the last term in Eq.~\eqref{eq:mass} and alters the fermionic mass term. Moreover, we observe in Eq.~\eqref{eq:H} that the chiral rotation yields an additional term in the final Hamiltonian. This is precisely the topological $\theta$-term, where $\theta \equiv 2\alpha$.

Let us now assume that the initial Hamiltonian $\hat{H}$ in Eq.~\eqref{eq:H} has a negative mass term. According to Eqs.~\eqref{eq:rot} and \eqref{eq:mass}, we can change the sign of this mass term with a chiral rotation by $\alpha = \theta/2 = \pi/2$, which yields $-m\to -m\exp(i\pi)=+m$. As discussed, this rotation gives rise to an additional term in the Hamiltonian,
\begin{align}
\hat{H}_{\theta = \pi} =\frac{\pi g^2}{16\pi^2} \sum_{\mu,\nu}\hat{F}^{\mu\nu}\hat{\tilde{F}}_{\mu\nu},
\label{eq:Hthetapi}
\end{align}
which is the topological $\theta$-term at $\theta = \pi$. Thus, we have shown that the theory with a negative mass term and without a topological $\theta$-term is equivalent to the theory with a positive mass term and the $\theta$-term at $\theta = \pi$. 
Note that the $\theta$-term in Eq.~\eqref{eq:Hthetapi} is a total derivative and therefore vanishes for gauge theories with a trivial topological vacuum structure, such as the non-compact U(1) gauge theory in (3+1)D. 

\section{Transfer matrix derivation of the extended Hamiltonian}
\label{sec:extended}

In this appendix, we derive the Hamiltonian corresponding to the extended U(1) lattice action~\cite{bhanot1982compact}
\begin{equation}
    S = -\frac{\beta}{2}\sum_{\substack{\vec{n},\mu,\nu;\\ \mu>\nu}}\left(U_{\vec{n},\mu \nu}+\text{h.c.}\right) - \frac{\gamma}{2}\sum_{\substack{\vec{n},\mu,\nu;\\ \mu>\nu}}\left(U_{\vec{n},\mu \nu}^{2}+\text{h.c.}\right),
    \label{eq:S_ext}
\end{equation}
where $\beta = 1/g^2$ and $\gamma$ are coupling constants. The U(1) Kogut-Susskind pure gauge Hamiltonian can be derived from the first sum using the transfer matrix method~\cite{creutz1977gauge}, as shown in Sec.~\ref{sec:lattice}. Here, we follow the same procedure to obtain the Hamiltonian corresponding to the second sum in Eq.~\eqref{eq:S_ext}.

In the temporal gauge, the matrix elements of the transfer operator~\eqref{eq:prodbasis} in the product basis is
\begin{multline}
    \bra{U'} \hat{T} \ket{U} = e^{\frac{a\gamma}{2a_0} \sum_{\vec{n},i} \left(U_{\vec{n},i}^{'2} U_{\vec{n},i}^{\dag 2} + U_{\vec{n},i}^{\dag'2} U_{\vec{n},i}^2\right)} \\
    \times e^{\frac{\gamma a_0}{2a} \sum_{\vec{n},j,k}\left(U_{\vec{n},jk}^2 + U_{\vec{n},jk}^{\dag 2}\right) }.
\end{multline}
In terms of the operators defined in Eqs.~\eqref{eq:usefulop1}--\eqref{eq:usefulop2}, we write the transfer operator as
\begin{align}
    \hat{T} &= \int_{g \in G} \prod_{\vec{n},i} dg_{\vec{n},i} \hat{R}_{\vec{n},i}(g_{\vec{n},i})e^{\frac{a \gamma}{2a_0}  \left(g_{\vec{n},i}^2 + g_{\vec{n},i}^{\dag 2}\right)} \nonumber \\
    & \quad \times e^{\frac{\gamma a_0}{2 a} \sum_{\vec{n},j,k}\left(\hat{U}_{\vec{n},jk}^2 + \hat{U}_{\vec{n},jk}^{\dag 2}\right) }\nonumber \\
    &= \int \prod_{\vec{n},i} dx_{\vec{n},i} e^{i x_{\vec{n},i} \hat{E}_{\vec{n},i}+\frac{a\gamma}{2a_0} 2\cos(2 x_{\vec{n},i}) } \nonumber \\
    & \quad \times e^{\frac{\gamma a_0}{2a} \sum_{\vec{n},j,k}\left(\hat{U}_{\vec{n},jk}^2 + \hat{U}_{\vec{n},jk}^{\dag 2}\right) }.
\end{align}
In the continuum limit, as $a_0 \rightarrow 0$, the integral is dominated by the maximum of $2\cos( 2 x_{\vec{n},i})$. Expanding around $x_{\vec{n},i} = 0$, where this maximum is located, we have
\begin{equation}
    2\cos( 2 x_{\vec{n},i}) \approx 2 - 4x_{\vec{n},i} x_{\vec{n},i}.
\end{equation}
Inserting the expansion into the integral, we obtain a Gaussian integral, which evaluates to
\begin{equation}
    \hat{T} \propto e^{-a_0\left[\frac{1}{8\gamma a}\sum_{\vec{n},j} \hat{E}_{\vec{n},j}^2-\frac{\gamma}{2 a} \sum_{\vec{n},j,k} \left(\hat{U}_{\vec{n},jk}^{2} + \text{h.c.}\right)\right]}.
\end{equation}
From this, we can directly read off the Hamiltonian corresponding to the second sum in Eq.~\eqref{eq:S_ext},
\begin{equation}
    \hat{H}_{\gamma} = \frac{1}{8\gamma a}\sum_{\vec{n}}\sum_{j=1}^{3} \hat{E}_{\vec{n},j}^2-\frac{\gamma}{2 a} \sum_{\vec{n}}\sum_{j,k=1;k> j}^{3} \left(\hat{U}_{\vec{n},jk}^{2} + \text{h.c.}\right).
\end{equation}

\bibliography{Papers}

\begin{thebibliography}{71}%
\makeatletter
\providecommand \@ifxundefined [1]{%
 \@ifx{#1\undefined}
}%
\providecommand \@ifnum [1]{%
 \ifnum #1\expandafter \@firstoftwo
 \else \expandafter \@secondoftwo
 \fi
}%
\providecommand \@ifx [1]{%
 \ifx #1\expandafter \@firstoftwo
 \else \expandafter \@secondoftwo
 \fi
}%
\providecommand \natexlab [1]{#1}%
\providecommand \enquote  [1]{``#1''}%
\providecommand \bibnamefont  [1]{#1}%
\providecommand \bibfnamefont [1]{#1}%
\providecommand \citenamefont [1]{#1}%
\providecommand \href@noop [0]{\@secondoftwo}%
\providecommand \href [0]{\begingroup \@sanitize@url \@href}%
\providecommand \@href[1]{\@@startlink{#1}\@@href}%
\providecommand \@@href[1]{\endgroup#1\@@endlink}%
\providecommand \@sanitize@url [0]{\catcode `\\12\catcode `\$12\catcode
  `\&12\catcode `\#12\catcode `\^12\catcode `\_12\catcode `\%12\relax}%
\providecommand \@@startlink[1]{}%
\providecommand \@@endlink[0]{}%
\providecommand \url  [0]{\begingroup\@sanitize@url \@url }%
\providecommand \@url [1]{\endgroup\@href {#1}{\urlprefix }}%
\providecommand \urlprefix  [0]{URL }%
\providecommand \Eprint [0]{\href }%
\providecommand \doibase [0]{http://dx.doi.org/}%
\providecommand \selectlanguage [0]{\@gobble}%
\providecommand \bibinfo  [0]{\@secondoftwo}%
\providecommand \bibfield  [0]{\@secondoftwo}%
\providecommand \translation [1]{[#1]}%
\providecommand \BibitemOpen [0]{}%
\providecommand \bibitemStop [0]{}%
\providecommand \bibitemNoStop [0]{.\EOS\space}%
\providecommand \EOS [0]{\spacefactor3000\relax}%
\providecommand \BibitemShut  [1]{\csname bibitem#1\endcsname}%
\let\auto@bib@innerbib\@empty
\bibitem [{\citenamefont {Aoki}\ \emph {et~al.}(2020)\citenamefont {Aoki} \emph
  {et~al.}}]{Aoki:2019cca}%
  \BibitemOpen
  \bibfield  {author} {\bibinfo {author} {\bibfnamefont {S.}~\bibnamefont
  {Aoki}} \emph {et~al.} (\bibinfo {collaboration} {Flavour Lattice Averaging
  Group}),\ }\bibfield  {title} {\enquote {\bibinfo {title} {{FLAG Review 2019:
  Flavour Lattice Averaging Group (FLAG)}},}\ }\href {\doibase
  10.1140/epjc/s10052-019-7354-7} {\bibfield  {journal} {\bibinfo  {journal}
  {Eur. Phys. J. C}\ }\textbf {\bibinfo {volume} {80}},\ \bibinfo {pages} {113}
  (\bibinfo {year} {2020})}\BibitemShut {NoStop}%
\bibitem [{\citenamefont {Troyer}\ and\ \citenamefont
  {Wiese}(2005)}]{Troyer:2004ge}%
  \BibitemOpen
  \bibfield  {author} {\bibinfo {author} {\bibfnamefont {M.}~\bibnamefont
  {Troyer}}\ and\ \bibinfo {author} {\bibfnamefont {U.-J.}\ \bibnamefont
  {Wiese}},\ }\bibfield  {title} {\enquote {\bibinfo {title} {{Computational
  complexity and fundamental limitations to fermionic quantum Monte Carlo
  simulations}},}\ }\href {\doibase 10.1103/PhysRevLett.94.170201} {\bibfield
  {journal} {\bibinfo  {journal} {Phys. Rev. Lett.}\ }\textbf {\bibinfo
  {volume} {94}},\ \bibinfo {pages} {170201} (\bibinfo {year}
  {2005})}\BibitemShut {NoStop}%
\bibitem [{\citenamefont {Gattringer}\ and\ \citenamefont
  {Langfeld}(2016)}]{Gattringer:2016kco}%
  \BibitemOpen
  \bibfield  {author} {\bibinfo {author} {\bibfnamefont {C.}~\bibnamefont
  {Gattringer}}\ and\ \bibinfo {author} {\bibfnamefont {K.}~\bibnamefont
  {Langfeld}},\ }\bibfield  {title} {\enquote {\bibinfo {title} {{Approaches to
  the sign problem in lattice field theory}},}\ }\href {\doibase
  10.1142/S0217751X16430077} {\bibfield  {journal} {\bibinfo  {journal} {Int.
  J. Mod. Phys. A}\ }\textbf {\bibinfo {volume} {31}},\ \bibinfo {pages}
  {1643007} (\bibinfo {year} {2016})}\BibitemShut {NoStop}%
\bibitem [{\citenamefont {Ba\~nuls}\ \emph {et~al.}(2018)\citenamefont
  {Ba\~nuls}, \citenamefont {Cichy}, \citenamefont {Cirac}, \citenamefont
  {Jansen},\ and\ \citenamefont {K\"uhn}}]{Banuls:2018jag}%
  \BibitemOpen
  \bibfield  {author} {\bibinfo {author} {\bibfnamefont {M.-C.}\ \bibnamefont
  {Ba\~nuls}}, \bibinfo {author} {\bibfnamefont {K.}~\bibnamefont {Cichy}},
  \bibinfo {author} {\bibfnamefont {J.~I.}\ \bibnamefont {Cirac}}, \bibinfo
  {author} {\bibfnamefont {K.}~\bibnamefont {Jansen}}, \ and\ \bibinfo {author}
  {\bibfnamefont {S.}~\bibnamefont {K\"uhn}},\ }\bibfield  {title} {\enquote
  {\bibinfo {title} {{Tensor Networks and their use for Lattice Gauge
  Theories}},}\ }\href {\doibase 10.22323/1.334.0022} {\bibfield  {journal}
  {\bibinfo  {journal} {PoS}\ }\textbf {\bibinfo {volume} {LATTICE2018}},\
  \bibinfo {pages} {022} (\bibinfo {year} {2018})}\BibitemShut {NoStop}%
\bibitem [{\citenamefont {Ba{\~{n}}uls}\ and\ \citenamefont
  {Cichy}(2020)}]{Banuls2019}%
  \BibitemOpen
  \bibfield  {author} {\bibinfo {author} {\bibfnamefont {M.-C.}\ \bibnamefont
  {Ba{\~{n}}uls}}\ and\ \bibinfo {author} {\bibfnamefont {K.}~\bibnamefont
  {Cichy}},\ }\bibfield  {title} {\enquote {\bibinfo {title} {Review on novel
  methods for lattice gauge theories},}\ }\href {\doibase
  10.1088/1361-6633/ab6311} {\bibfield  {journal} {\bibinfo  {journal} {Rep.
  Prog. Phys.}\ }\textbf {\bibinfo {volume} {83}},\ \bibinfo {pages} {024401}
  (\bibinfo {year} {2020})}\BibitemShut {NoStop}%
\bibitem [{\citenamefont {Kuramashi}\ and\ \citenamefont
  {Yoshimura}(2019)}]{Kuramashi:2018mmi}%
  \BibitemOpen
  \bibfield  {author} {\bibinfo {author} {\bibfnamefont {Y.}~\bibnamefont
  {Kuramashi}}\ and\ \bibinfo {author} {\bibfnamefont {Y.}~\bibnamefont
  {Yoshimura}},\ }\bibfield  {title} {\enquote {\bibinfo {title}
  {{Three-dimensional finite temperature Z$_{2}$ gauge theory with tensor
  network scheme}},}\ }\href {\doibase 10.1007/JHEP08(2019)023} {\bibfield
  {journal} {\bibinfo  {journal} {J. High Energy Phys.}\ }\textbf {\bibinfo
  {volume} {08}},\ \bibinfo {pages} {023} (\bibinfo {year} {2019})}\BibitemShut
  {NoStop}%
\bibitem [{\citenamefont {Felser}\ \emph {et~al.}(2020)\citenamefont {Felser},
  \citenamefont {Silvi}, \citenamefont {Collura},\ and\ \citenamefont
  {Montangero}}]{Felser2019}%
  \BibitemOpen
  \bibfield  {author} {\bibinfo {author} {\bibfnamefont {T.}~\bibnamefont
  {Felser}}, \bibinfo {author} {\bibfnamefont {P.}~\bibnamefont {Silvi}},
  \bibinfo {author} {\bibfnamefont {M.}~\bibnamefont {Collura}}, \ and\
  \bibinfo {author} {\bibfnamefont {S.}~\bibnamefont {Montangero}},\ }\bibfield
   {title} {\enquote {\bibinfo {title} {Two-dimensional quantum-link lattice
  quantum electrodynamics at finite density},}\ }\href {\doibase
  10.1103/PhysRevX.10.041040} {\bibfield  {journal} {\bibinfo  {journal} {Phys.
  Rev. X}\ }\textbf {\bibinfo {volume} {10}},\ \bibinfo {pages} {041040}
  (\bibinfo {year} {2020})}\BibitemShut {NoStop}%
\bibitem [{\citenamefont {Magnifico}\ \emph {et~al.}(2020)\citenamefont
  {Magnifico}, \citenamefont {Felser}, \citenamefont {Silvi},\ and\
  \citenamefont {Montangero}}]{Magnifico:2020bqt}%
  \BibitemOpen
  \bibfield  {author} {\bibinfo {author} {\bibfnamefont {G.}~\bibnamefont
  {Magnifico}}, \bibinfo {author} {\bibfnamefont {T.}~\bibnamefont {Felser}},
  \bibinfo {author} {\bibfnamefont {P.}~\bibnamefont {Silvi}}, \ and\ \bibinfo
  {author} {\bibfnamefont {S.}~\bibnamefont {Montangero}},\ }\bibfield  {title}
  {\enquote {\bibinfo {title} {{Lattice Quantum Electrodynamics in
  (3+1)-dimensions at finite density with Tensor Networks}},}\ }\href
  {https://arxiv.org/abs/2011.10658} {\bibfield  {journal} {\bibinfo  {journal}
  {arXiv preprint arXiv:2011.10658}\ } (\bibinfo {year} {2020})}\BibitemShut
  {NoStop}%
\bibitem [{\citenamefont {Ba\~nuls}\ \emph {et~al.}(2017)\citenamefont
  {Ba\~nuls}, \citenamefont {Cichy}, \citenamefont {Cirac}, \citenamefont
  {Jansen},\ and\ \citenamefont {K\"uhn}}]{Banuls2016a}%
  \BibitemOpen
  \bibfield  {author} {\bibinfo {author} {\bibfnamefont {M.-C.}\ \bibnamefont
  {Ba\~nuls}}, \bibinfo {author} {\bibfnamefont {K.}~\bibnamefont {Cichy}},
  \bibinfo {author} {\bibfnamefont {J.~I.}\ \bibnamefont {Cirac}}, \bibinfo
  {author} {\bibfnamefont {K.}~\bibnamefont {Jansen}}, \ and\ \bibinfo {author}
  {\bibfnamefont {S.}~\bibnamefont {K\"uhn}},\ }\bibfield  {title} {\enquote
  {\bibinfo {title} {Density induced phase transitions in the {Schwinger}
  model: A study with matrix product states},}\ }\href {\doibase
  10.1103/PhysRevLett.118.071601} {\bibfield  {journal} {\bibinfo  {journal}
  {Phys. Rev. Lett.}\ }\textbf {\bibinfo {volume} {118}},\ \bibinfo {pages}
  {071601} (\bibinfo {year} {2017})}\BibitemShut {NoStop}%
\bibitem [{\citenamefont {Silvi}\ \emph {et~al.}(2017)\citenamefont {Silvi},
  \citenamefont {Rico}, \citenamefont {Dalmonte}, \citenamefont {Tschirsich},\
  and\ \citenamefont {Montangero}}]{Silvi2016}%
  \BibitemOpen
  \bibfield  {author} {\bibinfo {author} {\bibfnamefont {P.}~\bibnamefont
  {Silvi}}, \bibinfo {author} {\bibfnamefont {E.}~\bibnamefont {Rico}},
  \bibinfo {author} {\bibfnamefont {M.}~\bibnamefont {Dalmonte}}, \bibinfo
  {author} {\bibfnamefont {F.}~\bibnamefont {Tschirsich}}, \ and\ \bibinfo
  {author} {\bibfnamefont {S.}~\bibnamefont {Montangero}},\ }\bibfield  {title}
  {\enquote {\bibinfo {title} {Finite-density phase diagram of a (1+1)-d
  non-abelian lattice gauge theory with tensor networks},}\ }\href {\doibase
  10.22331/q-2017-04-25-9} {\bibfield  {journal} {\bibinfo  {journal}
  {Quantum}\ }\textbf {\bibinfo {volume} {1}},\ \bibinfo {pages} {9} (\bibinfo
  {year} {2017})}\BibitemShut {NoStop}%
\bibitem [{\citenamefont {Byrnes}\ \emph {et~al.}(2002)\citenamefont {Byrnes},
  \citenamefont {Sriganesh}, \citenamefont {Bursill},\ and\ \citenamefont
  {Hamer}}]{Byrnes:2002gj}%
  \BibitemOpen
  \bibfield  {author} {\bibinfo {author} {\bibfnamefont {T.}~\bibnamefont
  {Byrnes}}, \bibinfo {author} {\bibfnamefont {P.}~\bibnamefont {Sriganesh}},
  \bibinfo {author} {\bibfnamefont {R.~J.}\ \bibnamefont {Bursill}}, \ and\
  \bibinfo {author} {\bibfnamefont {C.~J.}\ \bibnamefont {Hamer}},\ }\bibfield
  {title} {\enquote {\bibinfo {title} {{Density matrix renormalization group
  approach to the massive Schwinger model}},}\ }\href {\doibase
  10.1016/S0920-5632(02)01416-0} {\bibfield  {journal} {\bibinfo  {journal}
  {Nucl. Phys. B Proc. Suppl.}\ }\textbf {\bibinfo {volume} {109}},\ \bibinfo
  {pages} {202} (\bibinfo {year} {2002})}\BibitemShut {NoStop}%
\bibitem [{\citenamefont {Buyens}\ \emph {et~al.}(2017)\citenamefont {Buyens},
  \citenamefont {Montangero}, \citenamefont {Haegeman}, \citenamefont
  {Verstraete},\ and\ \citenamefont {Van~Acoleyen}}]{Buyens:2017crb}%
  \BibitemOpen
  \bibfield  {author} {\bibinfo {author} {\bibfnamefont {B.}~\bibnamefont
  {Buyens}}, \bibinfo {author} {\bibfnamefont {S.}~\bibnamefont {Montangero}},
  \bibinfo {author} {\bibfnamefont {J.}~\bibnamefont {Haegeman}}, \bibinfo
  {author} {\bibfnamefont {F.}~\bibnamefont {Verstraete}}, \ and\ \bibinfo
  {author} {\bibfnamefont {K.}~\bibnamefont {Van~Acoleyen}},\ }\bibfield
  {title} {\enquote {\bibinfo {title} {{Finite-representation approximation of
  lattice gauge theories at the continuum limit with tensor networks}},}\
  }\href {\doibase 10.1103/PhysRevD.95.094509} {\bibfield  {journal} {\bibinfo
  {journal} {Phys. Rev. D}\ }\textbf {\bibinfo {volume} {95}},\ \bibinfo
  {pages} {094509} (\bibinfo {year} {2017})}\BibitemShut {NoStop}%
\bibitem [{\citenamefont {Funcke}\ \emph {et~al.}(2020)\citenamefont {Funcke},
  \citenamefont {Jansen},\ and\ \citenamefont {K\"uhn}}]{Funcke:2019zna}%
  \BibitemOpen
  \bibfield  {author} {\bibinfo {author} {\bibfnamefont {L.}~\bibnamefont
  {Funcke}}, \bibinfo {author} {\bibfnamefont {K.}~\bibnamefont {Jansen}}, \
  and\ \bibinfo {author} {\bibfnamefont {S.}~\bibnamefont {K\"uhn}},\
  }\bibfield  {title} {\enquote {\bibinfo {title} {{Topological vacuum
  structure of the Schwinger model with matrix product states}},}\ }\href
  {\doibase 10.1103/PhysRevD.101.054507} {\bibfield  {journal} {\bibinfo
  {journal} {Phys. Rev. D}\ }\textbf {\bibinfo {volume} {101}},\ \bibinfo
  {pages} {054507} (\bibinfo {year} {2020})}\BibitemShut {NoStop}%
\bibitem [{\citenamefont {Martinez}\ \emph {et~al.}(2016)\citenamefont
  {Martinez} \emph {et~al.}}]{Martinez:2016yna}%
  \BibitemOpen
  \bibfield  {author} {\bibinfo {author} {\bibfnamefont {E.~A.}\ \bibnamefont
  {Martinez}} \emph {et~al.},\ }\bibfield  {title} {\enquote {\bibinfo {title}
  {{Real-time dynamics of lattice gauge theories with a few-qubit quantum
  computer}},}\ }\href {\doibase 10.1038/nature18318} {\bibfield  {journal}
  {\bibinfo  {journal} {Nature}\ }\textbf {\bibinfo {volume} {534}},\ \bibinfo
  {pages} {516} (\bibinfo {year} {2016})}\BibitemShut {NoStop}%
\bibitem [{\citenamefont {Kokail}\ \emph {et~al.}(2019)\citenamefont {Kokail}
  \emph {et~al.}}]{Kokail:2018eiw}%
  \BibitemOpen
  \bibfield  {author} {\bibinfo {author} {\bibfnamefont {C.}~\bibnamefont
  {Kokail}} \emph {et~al.},\ }\bibfield  {title} {\enquote {\bibinfo {title}
  {{Self-verifying variational quantum simulation of lattice models}},}\ }\href
  {\doibase 10.1038/s41586-019-1177-4} {\bibfield  {journal} {\bibinfo
  {journal} {Nature}\ }\textbf {\bibinfo {volume} {569}},\ \bibinfo {pages}
  {355} (\bibinfo {year} {2019})}\BibitemShut {NoStop}%
\bibitem [{\citenamefont {Klco}\ \emph {et~al.}(2018)\citenamefont {Klco},
  \citenamefont {Dumitrescu}, \citenamefont {McCaskey}, \citenamefont {Morris},
  \citenamefont {Pooser}, \citenamefont {Sanz}, \citenamefont {Solano},
  \citenamefont {Lougovski},\ and\ \citenamefont {Savage}}]{Klco:2018kyo}%
  \BibitemOpen
  \bibfield  {author} {\bibinfo {author} {\bibfnamefont {N.}~\bibnamefont
  {Klco}}, \bibinfo {author} {\bibfnamefont {E.~F.}\ \bibnamefont
  {Dumitrescu}}, \bibinfo {author} {\bibfnamefont {A.~J.}\ \bibnamefont
  {McCaskey}}, \bibinfo {author} {\bibfnamefont {T.~D.}\ \bibnamefont
  {Morris}}, \bibinfo {author} {\bibfnamefont {R.~C.}\ \bibnamefont {Pooser}},
  \bibinfo {author} {\bibfnamefont {M.}~\bibnamefont {Sanz}}, \bibinfo {author}
  {\bibfnamefont {E.}~\bibnamefont {Solano}}, \bibinfo {author} {\bibfnamefont
  {P.}~\bibnamefont {Lougovski}}, \ and\ \bibinfo {author} {\bibfnamefont
  {M.~J.}\ \bibnamefont {Savage}},\ }\bibfield  {title} {\enquote {\bibinfo
  {title} {{Quantum-classical computation of Schwinger model dynamics using
  quantum computers}},}\ }\href {\doibase 10.1103/PhysRevA.98.032331}
  {\bibfield  {journal} {\bibinfo  {journal} {Phys. Rev. A}\ }\textbf {\bibinfo
  {volume} {98}},\ \bibinfo {pages} {032331} (\bibinfo {year}
  {2018})}\BibitemShut {NoStop}%
\bibitem [{\citenamefont {Schweizer}\ \emph {et~al.}(2019)\citenamefont
  {Schweizer}, \citenamefont {Grusdt}, \citenamefont {Berngruber},
  \citenamefont {Barbiero}, \citenamefont {Demler}, \citenamefont {Goldman},
  \citenamefont {Bloch},\ and\ \citenamefont
  {Aidelsburger}}]{schweizer2019floquet}%
  \BibitemOpen
  \bibfield  {author} {\bibinfo {author} {\bibfnamefont {C.}~\bibnamefont
  {Schweizer}}, \bibinfo {author} {\bibfnamefont {F.}~\bibnamefont {Grusdt}},
  \bibinfo {author} {\bibfnamefont {M.}~\bibnamefont {Berngruber}}, \bibinfo
  {author} {\bibfnamefont {L.}~\bibnamefont {Barbiero}}, \bibinfo {author}
  {\bibfnamefont {E.}~\bibnamefont {Demler}}, \bibinfo {author} {\bibfnamefont
  {N.}~\bibnamefont {Goldman}}, \bibinfo {author} {\bibfnamefont
  {I.}~\bibnamefont {Bloch}}, \ and\ \bibinfo {author} {\bibfnamefont
  {M.}~\bibnamefont {Aidelsburger}},\ }\bibfield  {title} {\enquote {\bibinfo
  {title} {Floquet approach to {Z2} lattice gauge theories with ultracold atoms
  in optical lattices},}\ }\href {\doibase 10.1038/s41567-019-0649-7}
  {\bibfield  {journal} {\bibinfo  {journal} {Nature Phys.}\ }\textbf {\bibinfo
  {volume} {15}},\ \bibinfo {pages} {1168} (\bibinfo {year}
  {2019})}\BibitemShut {NoStop}%
\bibitem [{\citenamefont {G{\"o}rg}\ \emph {et~al.}(2019)\citenamefont
  {G{\"o}rg}, \citenamefont {Sandholzer}, \citenamefont {Minguzzi},
  \citenamefont {Desbuquois}, \citenamefont {Messer},\ and\ \citenamefont
  {Esslinger}}]{gorg2019realization}%
  \BibitemOpen
  \bibfield  {author} {\bibinfo {author} {\bibfnamefont {F.}~\bibnamefont
  {G{\"o}rg}}, \bibinfo {author} {\bibfnamefont {K.}~\bibnamefont
  {Sandholzer}}, \bibinfo {author} {\bibfnamefont {J.}~\bibnamefont
  {Minguzzi}}, \bibinfo {author} {\bibfnamefont {R.}~\bibnamefont
  {Desbuquois}}, \bibinfo {author} {\bibfnamefont {M.}~\bibnamefont {Messer}},
  \ and\ \bibinfo {author} {\bibfnamefont {T.}~\bibnamefont {Esslinger}},\
  }\bibfield  {title} {\enquote {\bibinfo {title} {Realization of
  density-dependent {Peierls} phases to engineer quantized gauge fields coupled
  to ultracold matter},}\ }\href {\doibase 10.1038/s41567-019-0615-4}
  {\bibfield  {journal} {\bibinfo  {journal} {Nature Phys.}\ }\textbf {\bibinfo
  {volume} {15}},\ \bibinfo {pages} {1161} (\bibinfo {year}
  {2019})}\BibitemShut {NoStop}%
\bibitem [{\citenamefont {Mil}\ \emph {et~al.}(2020)\citenamefont {Mil},
  \citenamefont {Zache}, \citenamefont {Hegde}, \citenamefont {Xia},
  \citenamefont {Bhatt}, \citenamefont {Oberthaler}, \citenamefont {Hauke},
  \citenamefont {Berges},\ and\ \citenamefont {Jendrzejewski}}]{Mil:2019pbt}%
  \BibitemOpen
  \bibfield  {author} {\bibinfo {author} {\bibfnamefont {A.}~\bibnamefont
  {Mil}}, \bibinfo {author} {\bibfnamefont {Torsten~V.}\ \bibnamefont {Zache}},
  \bibinfo {author} {\bibfnamefont {A.}~\bibnamefont {Hegde}}, \bibinfo
  {author} {\bibfnamefont {A.}~\bibnamefont {Xia}}, \bibinfo {author}
  {\bibfnamefont {R.~P.}\ \bibnamefont {Bhatt}}, \bibinfo {author}
  {\bibfnamefont {M.~K.}\ \bibnamefont {Oberthaler}}, \bibinfo {author}
  {\bibfnamefont {P.}~\bibnamefont {Hauke}}, \bibinfo {author} {\bibfnamefont
  {J.}~\bibnamefont {Berges}}, \ and\ \bibinfo {author} {\bibfnamefont
  {F.}~\bibnamefont {Jendrzejewski}},\ }\bibfield  {title} {\enquote {\bibinfo
  {title} {{A scalable realization of local U(1) gauge invariance in cold
  atomic mixtures}},}\ }\href {\doibase 10.1126/science.aaz5312} {\bibfield
  {journal} {\bibinfo  {journal} {Science}\ }\textbf {\bibinfo {volume}
  {367}},\ \bibinfo {pages} {1128} (\bibinfo {year} {2020})}\BibitemShut
  {NoStop}%
\bibitem [{\citenamefont {Klco}\ \emph {et~al.}(2020)\citenamefont {Klco},
  \citenamefont {Savage},\ and\ \citenamefont {Stryker}}]{klco20202}%
  \BibitemOpen
  \bibfield  {author} {\bibinfo {author} {\bibfnamefont {N.}~\bibnamefont
  {Klco}}, \bibinfo {author} {\bibfnamefont {M.~J.}\ \bibnamefont {Savage}}, \
  and\ \bibinfo {author} {\bibfnamefont {J.~R.}\ \bibnamefont {Stryker}},\
  }\bibfield  {title} {\enquote {\bibinfo {title} {{SU(2)} non-{Abelian} gauge
  field theory in one dimension on digital quantum computers},}\ }\href
  {\doibase 10.1103/PhysRevD.101.074512} {\bibfield  {journal} {\bibinfo
  {journal} {Phys. Rev. D}\ }\textbf {\bibinfo {volume} {101}},\ \bibinfo
  {pages} {074512} (\bibinfo {year} {2020})}\BibitemShut {NoStop}%
\bibitem [{\citenamefont {Yang}\ \emph {et~al.}(2020)\citenamefont {Yang},
  \citenamefont {Sun}, \citenamefont {Ott}, \citenamefont {Wang}, \citenamefont
  {Zache}, \citenamefont {Halimeh}, \citenamefont {Yuan}, \citenamefont
  {Hauke},\ and\ \citenamefont {Pan}}]{Yang2020}%
  \BibitemOpen
  \bibfield  {author} {\bibinfo {author} {\bibfnamefont {B.}~\bibnamefont
  {Yang}}, \bibinfo {author} {\bibfnamefont {H.}~\bibnamefont {Sun}}, \bibinfo
  {author} {\bibfnamefont {R.}~\bibnamefont {Ott}}, \bibinfo {author}
  {\bibfnamefont {H.-Y.}\ \bibnamefont {Wang}}, \bibinfo {author}
  {\bibfnamefont {T.~V.}\ \bibnamefont {Zache}}, \bibinfo {author}
  {\bibfnamefont {J.~C.}\ \bibnamefont {Halimeh}}, \bibinfo {author}
  {\bibfnamefont {Z.-S.}\ \bibnamefont {Yuan}}, \bibinfo {author}
  {\bibfnamefont {P.}~\bibnamefont {Hauke}}, \ and\ \bibinfo {author}
  {\bibfnamefont {J.-W.}\ \bibnamefont {Pan}},\ }\bibfield  {title} {\enquote
  {\bibinfo {title} {Observation of gauge invariance in a 71-site
  {Bose--Hubbard} quantum simulator},}\ }\href {\doibase
  10.1038/s41586-020-2910-8} {\bibfield  {journal} {\bibinfo  {journal}
  {Nature}\ }\textbf {\bibinfo {volume} {587}},\ \bibinfo {pages} {392}
  (\bibinfo {year} {2020})}\BibitemShut {NoStop}%
\bibitem [{\citenamefont {Atas}\ \emph {et~al.}(2021)\citenamefont {Atas},
  \citenamefont {Zhang}, \citenamefont {Lewis}, \citenamefont {Jahanpour},
  \citenamefont {Haase},\ and\ \citenamefont {Muschik}}]{atas20212}%
  \BibitemOpen
  \bibfield  {author} {\bibinfo {author} {\bibfnamefont {Y.}~\bibnamefont
  {Atas}}, \bibinfo {author} {\bibfnamefont {J.}~\bibnamefont {Zhang}},
  \bibinfo {author} {\bibfnamefont {R.}~\bibnamefont {Lewis}}, \bibinfo
  {author} {\bibfnamefont {A.}~\bibnamefont {Jahanpour}}, \bibinfo {author}
  {\bibfnamefont {J.~F.}\ \bibnamefont {Haase}}, \ and\ \bibinfo {author}
  {\bibfnamefont {C.~A.}\ \bibnamefont {Muschik}},\ }\bibfield  {title}
  {\enquote {\bibinfo {title} {{SU(2)} hadrons on a quantum computer},}\ }\href
  {https://arxiv.org/abs/2102.08920} {\bibfield  {journal} {\bibinfo  {journal}
  {arXiv preprint arXiv:2102.08920}\ } (\bibinfo {year} {2021})}\BibitemShut
  {NoStop}%
\bibitem [{\citenamefont {Ciavarella}\ \emph {et~al.}(2021)\citenamefont
  {Ciavarella}, \citenamefont {Klco},\ and\ \citenamefont
  {Savage}}]{ciavarella2021trailhead}%
  \BibitemOpen
  \bibfield  {author} {\bibinfo {author} {\bibfnamefont {A.}~\bibnamefont
  {Ciavarella}}, \bibinfo {author} {\bibfnamefont {N.}~\bibnamefont {Klco}}, \
  and\ \bibinfo {author} {\bibfnamefont {M.~J.}\ \bibnamefont {Savage}},\
  }\bibfield  {title} {\enquote {\bibinfo {title} {Trailhead for quantum
  simulation of {SU(3)} {Yang-Mills} lattice gauge theory in the local
  multiplet basis},}\ }\href {\doibase 10.1103/PhysRevD.103.094501} {\bibfield
  {journal} {\bibinfo  {journal} {Phys. Rev. D}\ }\textbf {\bibinfo {volume}
  {103}},\ \bibinfo {pages} {094501} (\bibinfo {year} {2021})}\BibitemShut
  {NoStop}%
\bibitem [{\citenamefont {Rahman}\ \emph {et~al.}(2021)\citenamefont {Rahman},
  \citenamefont {Lewis}, \citenamefont {Mendicelli},\ and\ \citenamefont
  {Powell}}]{rahman20212}%
  \BibitemOpen
  \bibfield  {author} {\bibinfo {author} {\bibfnamefont {S.~A.}\ \bibnamefont
  {Rahman}}, \bibinfo {author} {\bibfnamefont {R.}~\bibnamefont {Lewis}},
  \bibinfo {author} {\bibfnamefont {E.}~\bibnamefont {Mendicelli}}, \ and\
  \bibinfo {author} {\bibfnamefont {S.}~\bibnamefont {Powell}},\ }\bibfield
  {title} {\enquote {\bibinfo {title} {{SU(2) lattice gauge theory on a quantum
  annealer}},}\ }\href {https://arxiv.org/abs/2103.08661} {\bibfield  {journal}
  {\bibinfo  {journal} {arXiv preprint arXiv:2103.08661}\ } (\bibinfo {year}
  {2021})}\BibitemShut {NoStop}%
\bibitem [{\citenamefont {Gonz{\'a}lez-Cuadra}\ \emph
  {et~al.}(2017)\citenamefont {Gonz{\'a}lez-Cuadra}, \citenamefont {Zohar},\
  and\ \citenamefont {Cirac}}]{gonzalez2017quantum}%
  \BibitemOpen
  \bibfield  {author} {\bibinfo {author} {\bibfnamefont {D.}~\bibnamefont
  {Gonz{\'a}lez-Cuadra}}, \bibinfo {author} {\bibfnamefont {E.}~\bibnamefont
  {Zohar}}, \ and\ \bibinfo {author} {\bibfnamefont {J.~I.}\ \bibnamefont
  {Cirac}},\ }\bibfield  {title} {\enquote {\bibinfo {title} {{Quantum
  simulation of the Abelian-Higgs lattice gauge theory with ultracold
  atoms}},}\ }\href {\doibase 10.1088/1367-2630/aa6f37} {\bibfield  {journal}
  {\bibinfo  {journal} {New J. Phys.}\ }\textbf {\bibinfo {volume} {19}},\
  \bibinfo {pages} {063038} (\bibinfo {year} {2017})}\BibitemShut {NoStop}%
\bibitem [{\citenamefont {Bender}\ \emph {et~al.}(2018)\citenamefont {Bender},
  \citenamefont {Zohar}, \citenamefont {Farace},\ and\ \citenamefont
  {Cirac}}]{bender2018digital}%
  \BibitemOpen
  \bibfield  {author} {\bibinfo {author} {\bibfnamefont {J.}~\bibnamefont
  {Bender}}, \bibinfo {author} {\bibfnamefont {E.}~\bibnamefont {Zohar}},
  \bibinfo {author} {\bibfnamefont {A.}~\bibnamefont {Farace}}, \ and\ \bibinfo
  {author} {\bibfnamefont {J.~I.}\ \bibnamefont {Cirac}},\ }\bibfield  {title}
  {\enquote {\bibinfo {title} {Digital quantum simulation of lattice gauge
  theories in three spatial dimensions},}\ }\href {\doibase
  10.1088/1367-2630/aadb71} {\bibfield  {journal} {\bibinfo  {journal} {New J.
  Phys.}\ }\textbf {\bibinfo {volume} {20}},\ \bibinfo {pages} {093001}
  (\bibinfo {year} {2018})}\BibitemShut {NoStop}%
\bibitem [{\citenamefont {Banuls}\ \emph {et~al.}(2020)\citenamefont {Banuls}
  \emph {et~al.}}]{banuls2020simulating}%
  \BibitemOpen
  \bibfield  {author} {\bibinfo {author} {\bibfnamefont {M.-C.}\ \bibnamefont
  {Banuls}} \emph {et~al.},\ }\bibfield  {title} {\enquote {\bibinfo {title}
  {Simulating lattice gauge theories within quantum technologies},}\ }\href
  {\doibase 10.1140/epjd/e2020-100571-8} {\bibfield  {journal} {\bibinfo
  {journal} {Eur. Phys. J. D}\ }\textbf {\bibinfo {volume} {74}},\ \bibinfo
  {pages} {165} (\bibinfo {year} {2020})}\BibitemShut {NoStop}%
\bibitem [{\citenamefont {Crewther}\ \emph {et~al.}(1979)\citenamefont
  {Crewther}, \citenamefont {Vecchia}, \citenamefont {Veneziano},\ and\
  \citenamefont {Witten}}]{Crewther:1979}%
  \BibitemOpen
  \bibfield  {author} {\bibinfo {author} {\bibfnamefont {R.~J.}\ \bibnamefont
  {Crewther}}, \bibinfo {author} {\bibfnamefont {P.~Di}\ \bibnamefont
  {Vecchia}}, \bibinfo {author} {\bibfnamefont {G.}~\bibnamefont {Veneziano}},
  \ and\ \bibinfo {author} {\bibfnamefont {E.}~\bibnamefont {Witten}},\
  }\bibfield  {title} {\enquote {\bibinfo {title} {Chiral estimate of the
  electric dipole moment of the neutron in quantum chromodynamics},}\ }\href
  {\doibase 10.1016/0370-2693(79)90128-X} {\bibfield  {journal} {\bibinfo
  {journal} {Phys. Lett. B}\ }\textbf {\bibinfo {volume} {88}},\ \bibinfo
  {pages} {123} (\bibinfo {year} {1979})}\BibitemShut {NoStop}%
\bibitem [{\citenamefont {Abel}\ \emph {et~al.}(2020)\citenamefont {Abel} \emph
  {et~al.}}]{Abel:2020}%
  \BibitemOpen
  \bibfield  {author} {\bibinfo {author} {\bibfnamefont {C.}~\bibnamefont
  {Abel}} \emph {et~al.},\ }\bibfield  {title} {\enquote {\bibinfo {title}
  {Measurement of the permanent electric dipole moment of the neutron},}\
  }\href {\doibase 10.1103/PhysRevLett.124.081803} {\bibfield  {journal}
  {\bibinfo  {journal} {Phys. Rev. Lett.}\ }\textbf {\bibinfo {volume} {124}},\
  \bibinfo {pages} {081803} (\bibinfo {year} {2020})}\BibitemShut {NoStop}%
\bibitem [{\citenamefont {Hook}(2018)}]{Hook:2018dlk}%
  \BibitemOpen
  \bibfield  {author} {\bibinfo {author} {\bibfnamefont {A.}~\bibnamefont
  {Hook}},\ }\bibfield  {title} {\enquote {\bibinfo {title} {{TASI lectures on
  the strong CP problem and axions}},}\ }\href
  {https://arxiv.org/abs/1812.02669} {\bibfield  {journal} {\bibinfo  {journal}
  {arXiv preprint arXiv:1812.02669}\ } (\bibinfo {year} {2018})}\BibitemShut
  {NoStop}%
\bibitem [{\citenamefont {Borsanyi}\ \emph {et~al.}(2016)\citenamefont
  {Borsanyi} \emph {et~al.}}]{Borsanyi:2016ksw}%
  \BibitemOpen
  \bibfield  {author} {\bibinfo {author} {\bibfnamefont {S.}~\bibnamefont
  {Borsanyi}} \emph {et~al.},\ }\bibfield  {title} {\enquote {\bibinfo {title}
  {{Calculation of the axion mass based on high-temperature lattice quantum
  chromodynamics}},}\ }\href {\doibase 10.1038/nature20115} {\bibfield
  {journal} {\bibinfo  {journal} {Nature}\ }\textbf {\bibinfo {volume} {539}},\
  \bibinfo {pages} {69} (\bibinfo {year} {2016})}\BibitemShut {NoStop}%
\bibitem [{\citenamefont {Dragos}\ \emph {et~al.}(2021)\citenamefont {Dragos},
  \citenamefont {Luu}, \citenamefont {Shindler}, \citenamefont {de~Vries},\
  and\ \citenamefont {Yousif}}]{Dragos:2019oxn}%
  \BibitemOpen
  \bibfield  {author} {\bibinfo {author} {\bibfnamefont {J.}~\bibnamefont
  {Dragos}}, \bibinfo {author} {\bibfnamefont {T.}~\bibnamefont {Luu}},
  \bibinfo {author} {\bibfnamefont {A.}~\bibnamefont {Shindler}}, \bibinfo
  {author} {\bibfnamefont {J.}~\bibnamefont {de~Vries}}, \ and\ \bibinfo
  {author} {\bibfnamefont {A.}~\bibnamefont {Yousif}},\ }\bibfield  {title}
  {\enquote {\bibinfo {title} {{Confirming the Existence of the strong CP
  Problem in Lattice QCD with the Gradient Flow}},}\ }\href {\doibase
  10.1103/PhysRevC.103.015202} {\bibfield  {journal} {\bibinfo  {journal}
  {Phys. Rev. C}\ }\textbf {\bibinfo {volume} {103}},\ \bibinfo {pages}
  {015202} (\bibinfo {year} {2021})}\BibitemShut {NoStop}%
\bibitem [{\citenamefont {Alexandrou}\ \emph
  {et~al.}(2020{\natexlab{a}})\citenamefont {Alexandrou}, \citenamefont
  {Finkenrath}, \citenamefont {Funcke}, \citenamefont {Jansen}, \citenamefont
  {Kostrzewa}, \citenamefont {Pittler},\ and\ \citenamefont
  {Urbach}}]{Alexandrou:2020bkd}%
  \BibitemOpen
  \bibfield  {author} {\bibinfo {author} {\bibfnamefont {C.}~\bibnamefont
  {Alexandrou}}, \bibinfo {author} {\bibfnamefont {J.}~\bibnamefont
  {Finkenrath}}, \bibinfo {author} {\bibfnamefont {L.}~\bibnamefont {Funcke}},
  \bibinfo {author} {\bibfnamefont {K.}~\bibnamefont {Jansen}}, \bibinfo
  {author} {\bibfnamefont {B.}~\bibnamefont {Kostrzewa}}, \bibinfo {author}
  {\bibfnamefont {F.}~\bibnamefont {Pittler}}, \ and\ \bibinfo {author}
  {\bibfnamefont {C.}~\bibnamefont {Urbach}},\ }\bibfield  {title} {\enquote
  {\bibinfo {title} {{Ruling Out the Massless Up-Quark Solution to the Strong
  $\pmb{CP}$ Problem by Computing the Topological Mass Contribution with
  Lattice QCD}},}\ }\href {\doibase 10.1103/PhysRevLett.125.232001} {\bibfield
  {journal} {\bibinfo  {journal} {Phys. Rev. Lett.}\ }\textbf {\bibinfo
  {volume} {125}},\ \bibinfo {pages} {232001} (\bibinfo {year}
  {2020}{\natexlab{a}})}\BibitemShut {NoStop}%
\bibitem [{\citenamefont {Coleman}(1976)}]{Coleman:1976uz}%
  \BibitemOpen
  \bibfield  {author} {\bibinfo {author} {\bibfnamefont {S.~R.}\ \bibnamefont
  {Coleman}},\ }\bibfield  {title} {\enquote {\bibinfo {title} {{More About the
  Massive Schwinger Model}},}\ }\href {\doibase 10.1016/0003-4916(76)90280-3}
  {\bibfield  {journal} {\bibinfo  {journal} {Annals Phys.}\ }\textbf {\bibinfo
  {volume} {101}},\ \bibinfo {pages} {239} (\bibinfo {year}
  {1976})}\BibitemShut {NoStop}%
\bibitem [{\citenamefont {Deser}\ \emph
  {et~al.}(1982{\natexlab{a}})\citenamefont {Deser}, \citenamefont {Jackiw},\
  and\ \citenamefont {Templeton}}]{Deser:1982vy}%
  \BibitemOpen
  \bibfield  {author} {\bibinfo {author} {\bibfnamefont {S.}~\bibnamefont
  {Deser}}, \bibinfo {author} {\bibfnamefont {R.}~\bibnamefont {Jackiw}}, \
  and\ \bibinfo {author} {\bibfnamefont {S.}~\bibnamefont {Templeton}},\
  }\bibfield  {title} {\enquote {\bibinfo {title} {{Three-Dimensional Massive
  Gauge Theories}},}\ }\href {\doibase 10.1103/PhysRevLett.48.975} {\bibfield
  {journal} {\bibinfo  {journal} {Phys. Rev. Lett.}\ }\textbf {\bibinfo
  {volume} {48}},\ \bibinfo {pages} {975} (\bibinfo {year}
  {1982}{\natexlab{a}})}\BibitemShut {NoStop}%
\bibitem [{\citenamefont {Deser}\ \emph
  {et~al.}(1982{\natexlab{b}})\citenamefont {Deser}, \citenamefont {Jackiw},\
  and\ \citenamefont {Templeton}}]{Deser:1981wh}%
  \BibitemOpen
  \bibfield  {author} {\bibinfo {author} {\bibfnamefont {S.}~\bibnamefont
  {Deser}}, \bibinfo {author} {\bibfnamefont {R.}~\bibnamefont {Jackiw}}, \
  and\ \bibinfo {author} {\bibfnamefont {S.}~\bibnamefont {Templeton}},\
  }\bibfield  {title} {\enquote {\bibinfo {title} {{Topologically Massive Gauge
  Theories}},}\ }\href {\doibase 10.1016/0003-4916(82)90164-6} {\bibfield
  {journal} {\bibinfo  {journal} {Annals Phys.}\ }\textbf {\bibinfo {volume}
  {140}},\ \bibinfo {pages} {372} (\bibinfo {year} {1982}{\natexlab{b}})},\
  \bibinfo {note} {[Erratum: Annals Phys. 185, 406 (1988)]}\BibitemShut
  {NoStop}%
\bibitem [{\citenamefont {Witten}(2016)}]{Witten:2015aoa}%
  \BibitemOpen
  \bibfield  {author} {\bibinfo {author} {\bibfnamefont {E.}~\bibnamefont
  {Witten}},\ }\bibfield  {title} {\enquote {\bibinfo {title} {{Three lectures
  on topological phases of matter}},}\ }\href {\doibase
  10.1393/ncr/i2016-10125-3} {\bibfield  {journal} {\bibinfo  {journal} {Riv.
  Nuovo Cim.}\ }\textbf {\bibinfo {volume} {39}},\ \bibinfo {pages} {313}
  (\bibinfo {year} {2016})}\BibitemShut {NoStop}%
\bibitem [{\citenamefont {Alexandrou}\ \emph
  {et~al.}(2020{\natexlab{b}})\citenamefont {Alexandrou}, \citenamefont
  {Athenodorou}, \citenamefont {Cichy}, \citenamefont {Dromard}, \citenamefont
  {Garcia-Ramos}, \citenamefont {Jansen}, \citenamefont {Wenger},\ and\
  \citenamefont {Zimmermann}}]{Alexandrou:2017hqw}%
  \BibitemOpen
  \bibfield  {author} {\bibinfo {author} {\bibfnamefont {C.}~\bibnamefont
  {Alexandrou}}, \bibinfo {author} {\bibfnamefont {A.}~\bibnamefont
  {Athenodorou}}, \bibinfo {author} {\bibfnamefont {K.}~\bibnamefont {Cichy}},
  \bibinfo {author} {\bibfnamefont {A.}~\bibnamefont {Dromard}}, \bibinfo
  {author} {\bibfnamefont {E.}~\bibnamefont {Garcia-Ramos}}, \bibinfo {author}
  {\bibfnamefont {K.}~\bibnamefont {Jansen}}, \bibinfo {author} {\bibfnamefont
  {U.}~\bibnamefont {Wenger}}, \ and\ \bibinfo {author} {\bibfnamefont
  {F.}~\bibnamefont {Zimmermann}},\ }\bibfield  {title} {\enquote {\bibinfo
  {title} {{Comparison of topological charge definitions in Lattice QCD}},}\
  }\href {\doibase 10.1140/epjc/s10052-020-7984-9} {\bibfield  {journal}
  {\bibinfo  {journal} {Eur. Phys. J. C}\ }\textbf {\bibinfo {volume} {80}},\
  \bibinfo {pages} {424} (\bibinfo {year} {2020}{\natexlab{b}})}\BibitemShut
  {NoStop}%
\bibitem [{\citenamefont {Creutz}(1977)}]{creutz1977gauge}%
  \BibitemOpen
  \bibfield  {author} {\bibinfo {author} {\bibfnamefont {M.}~\bibnamefont
  {Creutz}},\ }\bibfield  {title} {\enquote {\bibinfo {title} {Gauge fixing,
  the transfer matrix, and confinement on a lattice},}\ }\href {\doibase
  10.1103/PhysRevD.15.1128} {\bibfield  {journal} {\bibinfo  {journal} {Phys.
  Rev. D}\ }\textbf {\bibinfo {volume} {15}},\ \bibinfo {pages} {1128}
  (\bibinfo {year} {1977})}\BibitemShut {NoStop}%
\bibitem [{\citenamefont {Di~Vecchia}\ \emph {et~al.}(1981)\citenamefont
  {Di~Vecchia}, \citenamefont {Fabricius}, \citenamefont {Rossi},\ and\
  \citenamefont {Veneziano}}]{DiVecchia:1981aev}%
  \BibitemOpen
  \bibfield  {author} {\bibinfo {author} {\bibfnamefont {P.}~\bibnamefont
  {Di~Vecchia}}, \bibinfo {author} {\bibfnamefont {K.}~\bibnamefont
  {Fabricius}}, \bibinfo {author} {\bibfnamefont {G.~C.}\ \bibnamefont
  {Rossi}}, \ and\ \bibinfo {author} {\bibfnamefont {G.}~\bibnamefont
  {Veneziano}},\ }\bibfield  {title} {\enquote {\bibinfo {title} {{Preliminary
  Evidence for U$_A(1)$ Breaking in QCD from Lattice Calculations}},}\ }\href
  {\doibase 10.1016/0550-3213(81)90432-6} {\bibfield  {journal} {\bibinfo
  {journal} {Nucl. Phys. B}\ }\textbf {\bibinfo {volume} {192}},\ \bibinfo
  {pages} {392} (\bibinfo {year} {1981})}\BibitemShut {NoStop}%
\bibitem [{\citenamefont {Cardy}\ and\ \citenamefont
  {Rabinovici}(1982)}]{Cardy:1981qy}%
  \BibitemOpen
  \bibfield  {author} {\bibinfo {author} {\bibfnamefont {J.~L.}\ \bibnamefont
  {Cardy}}\ and\ \bibinfo {author} {\bibfnamefont {E.}~\bibnamefont
  {Rabinovici}},\ }\bibfield  {title} {\enquote {\bibinfo {title} {{Phase
  Structure of Z(p) Models in the Presence of a Theta Parameter}},}\ }\href
  {\doibase 10.1016/0550-3213(82)90463-1} {\bibfield  {journal} {\bibinfo
  {journal} {Nucl. Phys. B}\ }\textbf {\bibinfo {volume} {205}},\ \bibinfo
  {pages} {1} (\bibinfo {year} {1982})}\BibitemShut {NoStop}%
\bibitem [{\citenamefont {Cardy}(1982)}]{cardy1982duality}%
  \BibitemOpen
  \bibfield  {author} {\bibinfo {author} {\bibfnamefont {J.~L.}\ \bibnamefont
  {Cardy}},\ }\bibfield  {title} {\enquote {\bibinfo {title} {Duality and the
  $\theta$ parameter in abelian lattice models},}\ }\href {\doibase
  10.1016/0550-3213(82)90464-3} {\bibfield  {journal} {\bibinfo  {journal}
  {Nucl. Phys. B}\ }\textbf {\bibinfo {volume} {205}},\ \bibinfo {pages} {17}
  (\bibinfo {year} {1982})}\BibitemShut {NoStop}%
\bibitem [{\citenamefont {Honda}\ and\ \citenamefont
  {Tanizaki}(2020)}]{honda2020topological}%
  \BibitemOpen
  \bibfield  {author} {\bibinfo {author} {\bibfnamefont {M.}~\bibnamefont
  {Honda}}\ and\ \bibinfo {author} {\bibfnamefont {Y.}~\bibnamefont
  {Tanizaki}},\ }\bibfield  {title} {\enquote {\bibinfo {title} {Topological
  aspects of 4d abelian lattice gauge theories with the $\theta$ parameter},}\
  }\href {\doibase 10.1007/JHEP12(2020)154} {\bibfield  {journal} {\bibinfo
  {journal} {J. High Energy Phys.}\ }\textbf {\bibinfo {volume} {2020}},\
  \bibinfo {pages} {1} (\bibinfo {year} {2020})}\BibitemShut {NoStop}%
\bibitem [{\citenamefont {Dashen}(1971)}]{Dashen1971}%
  \BibitemOpen
  \bibfield  {author} {\bibinfo {author} {\bibfnamefont {R.}~\bibnamefont
  {Dashen}},\ }\bibfield  {title} {\enquote {\bibinfo {title} {Some features of
  chiral symmetry breaking},}\ }\href {\doibase 10.1103/PhysRevD.3.1879}
  {\bibfield  {journal} {\bibinfo  {journal} {Phys. Rev. D}\ }\textbf {\bibinfo
  {volume} {3}},\ \bibinfo {pages} {1879} (\bibinfo {year} {1971})}\BibitemShut
  {NoStop}%
\bibitem [{\citenamefont {Di~Vecchia}\ and\ \citenamefont
  {Veneziano}(1980)}]{DiVecchia:1980yfw}%
  \BibitemOpen
  \bibfield  {author} {\bibinfo {author} {\bibfnamefont {P.}~\bibnamefont
  {Di~Vecchia}}\ and\ \bibinfo {author} {\bibfnamefont {G.}~\bibnamefont
  {Veneziano}},\ }\bibfield  {title} {\enquote {\bibinfo {title} {{Chiral
  Dynamics in the Large N Limit}},}\ }\href {\doibase
  10.1016/0550-3213(80)90370-3} {\bibfield  {journal} {\bibinfo  {journal}
  {Nucl. Phys. B}\ }\textbf {\bibinfo {volume} {171}},\ \bibinfo {pages} {253}
  (\bibinfo {year} {1980})}\BibitemShut {NoStop}%
\bibitem [{\citenamefont {Witten}(1980)}]{Witten:1980sp}%
  \BibitemOpen
  \bibfield  {author} {\bibinfo {author} {\bibfnamefont {E.}~\bibnamefont
  {Witten}},\ }\bibfield  {title} {\enquote {\bibinfo {title} {{Large N Chiral
  Dynamics}},}\ }\href {\doibase 10.1016/0003-4916(80)90325-5} {\bibfield
  {journal} {\bibinfo  {journal} {Annals Phys.}\ }\textbf {\bibinfo {volume}
  {128}},\ \bibinfo {pages} {363} (\bibinfo {year} {1980})}\BibitemShut
  {NoStop}%
\bibitem [{\citenamefont {Haase}\ \emph {et~al.}(2021)\citenamefont {Haase},
  \citenamefont {Dellantonio}, \citenamefont {Celi}, \citenamefont {Paulson},
  \citenamefont {Kan}, \citenamefont {Jansen},\ and\ \citenamefont
  {Muschik}}]{haase2021resource}%
  \BibitemOpen
  \bibfield  {author} {\bibinfo {author} {\bibfnamefont {J.~F.}\ \bibnamefont
  {Haase}}, \bibinfo {author} {\bibfnamefont {L.}~\bibnamefont {Dellantonio}},
  \bibinfo {author} {\bibfnamefont {A.}~\bibnamefont {Celi}}, \bibinfo {author}
  {\bibfnamefont {D.}~\bibnamefont {Paulson}}, \bibinfo {author} {\bibfnamefont
  {A.}~\bibnamefont {Kan}}, \bibinfo {author} {\bibfnamefont {K.}~\bibnamefont
  {Jansen}}, \ and\ \bibinfo {author} {\bibfnamefont {C.~A.}\ \bibnamefont
  {Muschik}},\ }\bibfield  {title} {\enquote {\bibinfo {title} {A resource
  efficient approach for quantum and classical simulations of gauge theories in
  particle physics},}\ }\href {\doibase 10.22331/q-2021-02-04-393} {\bibfield
  {journal} {\bibinfo  {journal} {Quantum}\ }\textbf {\bibinfo {volume} {5}},\
  \bibinfo {pages} {393} (\bibinfo {year} {2021})}\BibitemShut {NoStop}%
\bibitem [{\citenamefont {Wilson}(1974)}]{wilson1974confinement}%
  \BibitemOpen
  \bibfield  {author} {\bibinfo {author} {\bibfnamefont {K.~G.}\ \bibnamefont
  {Wilson}},\ }\bibfield  {title} {\enquote {\bibinfo {title} {Confinement of
  quarks},}\ }\href {\doibase 10.1103/PhysRevD.10.2445} {\bibfield  {journal}
  {\bibinfo  {journal} {Phys. Rev. D}\ }\textbf {\bibinfo {volume} {10}},\
  \bibinfo {pages} {2445} (\bibinfo {year} {1974})}\BibitemShut {NoStop}%
\bibitem [{\citenamefont {Kogut}\ and\ \citenamefont
  {Susskind}(1975)}]{kogut1975hamiltonian}%
  \BibitemOpen
  \bibfield  {author} {\bibinfo {author} {\bibfnamefont {J.}~\bibnamefont
  {Kogut}}\ and\ \bibinfo {author} {\bibfnamefont {L.}~\bibnamefont
  {Susskind}},\ }\bibfield  {title} {\enquote {\bibinfo {title} {Hamiltonian
  formulation of {Wilson's} lattice gauge theories},}\ }\href {\doibase
  10.1103/PhysRevD.11.395} {\bibfield  {journal} {\bibinfo  {journal} {Phys.
  Rev. D}\ }\textbf {\bibinfo {volume} {11}},\ \bibinfo {pages} {395} (\bibinfo
  {year} {1975})}\BibitemShut {NoStop}%
\bibitem [{\citenamefont {Rothe}(2005)}]{rothe2005lattice}%
  \BibitemOpen
  \bibfield  {author} {\bibinfo {author} {\bibfnamefont {H.~J.}\ \bibnamefont
  {Rothe}},\ }\href@noop {} {\emph {\bibinfo {title} {Lattice Gauge Theories:
  An Introduction Third Edition}}},\ Vol.~\bibinfo {volume} {74}\ (\bibinfo
  {publisher} {World Scientific Publishing Company},\ \bibinfo {year}
  {2005})\BibitemShut {NoStop}%
\bibitem [{\citenamefont {Peskin}(1978)}]{Peskin1978}%
  \BibitemOpen
  \bibfield  {author} {\bibinfo {author} {\bibfnamefont {M.}~\bibnamefont
  {Peskin}},\ }\href@noop {} {\enquote {\bibinfo {title} {{Cornell University
  preprint CLNS 395, Thesis}},}\ } (\bibinfo {year} {1978})\BibitemShut
  {NoStop}%
\bibitem [{\citenamefont {Kaplan}\ and\ \citenamefont
  {Stryker}(2020)}]{PhysRevD.102.094515}%
  \BibitemOpen
  \bibfield  {author} {\bibinfo {author} {\bibfnamefont {David~B.}\
  \bibnamefont {Kaplan}}\ and\ \bibinfo {author} {\bibfnamefont {Jesse~R.}\
  \bibnamefont {Stryker}},\ }\bibfield  {title} {\enquote {\bibinfo {title}
  {Gauss's law, duality, and the hamiltonian formulation of u(1) lattice gauge
  theory},}\ }\href {\doibase 10.1103/PhysRevD.102.094515} {\bibfield
  {journal} {\bibinfo  {journal} {Phys. Rev. D}\ }\textbf {\bibinfo {volume}
  {102}},\ \bibinfo {pages} {094515} (\bibinfo {year} {2020})}\BibitemShut
  {NoStop}%
\bibitem [{\citenamefont {Bhanot}\ \emph {et~al.}(1984)\citenamefont {Bhanot},
  \citenamefont {Rabinovici}, \citenamefont {Seiberg},\ and\ \citenamefont
  {Woit}}]{bhanot1984lattice}%
  \BibitemOpen
  \bibfield  {author} {\bibinfo {author} {\bibfnamefont {G.}~\bibnamefont
  {Bhanot}}, \bibinfo {author} {\bibfnamefont {E.}~\bibnamefont {Rabinovici}},
  \bibinfo {author} {\bibfnamefont {N.}~\bibnamefont {Seiberg}}, \ and\
  \bibinfo {author} {\bibfnamefont {P.}~\bibnamefont {Woit}},\ }\bibfield
  {title} {\enquote {\bibinfo {title} {Lattice $\theta$ vacua},}\ }\href
  {\doibase 10.1016/0550-3213(84)90214-1} {\bibfield  {journal} {\bibinfo
  {journal} {Nucl. Phys. B}\ }\textbf {\bibinfo {volume} {230}},\ \bibinfo
  {pages} {291} (\bibinfo {year} {1984})}\BibitemShut {NoStop}%
\bibitem [{\citenamefont {Bilal}(2008)}]{bilal2008lectures}%
  \BibitemOpen
  \bibfield  {author} {\bibinfo {author} {\bibfnamefont {A.}~\bibnamefont
  {Bilal}},\ }\bibfield  {title} {\enquote {\bibinfo {title} {Lectures on
  anomalies},}\ }\href {https://arxiv.org/abs/0802.0634v1} {\bibfield
  {journal} {\bibinfo  {journal} {arXiv preprint arXiv:0802.0634}\ } (\bibinfo
  {year} {2008})}\BibitemShut {NoStop}%
\bibitem [{\citenamefont {Cohen}\ \emph {et~al.}(2021)\citenamefont {Cohen},
  \citenamefont {Lamm}, \citenamefont {Lawrence},\ and\ \citenamefont
  {Yamauchi}}]{cohen2021quantum}%
  \BibitemOpen
  \bibfield  {author} {\bibinfo {author} {\bibfnamefont {T.~D.}\ \bibnamefont
  {Cohen}}, \bibinfo {author} {\bibfnamefont {H.}~\bibnamefont {Lamm}},
  \bibinfo {author} {\bibfnamefont {S.}~\bibnamefont {Lawrence}}, \ and\
  \bibinfo {author} {\bibfnamefont {Y.}~\bibnamefont {Yamauchi}},\ }\bibfield
  {title} {\enquote {\bibinfo {title} {Quantum algorithms for transport
  coefficients in gauge theories},}\ }\href {https://arxiv.org/abs/2104.02024}
  {\bibfield  {journal} {\bibinfo  {journal} {arXiv preprint arXiv:2104.02024}\
  } (\bibinfo {year} {2021})}\BibitemShut {NoStop}%
\bibitem [{\citenamefont {Zohar}\ and\ \citenamefont
  {Burrello}(2015)}]{zohar2015formulation}%
  \BibitemOpen
  \bibfield  {author} {\bibinfo {author} {\bibfnamefont {E.}~\bibnamefont
  {Zohar}}\ and\ \bibinfo {author} {\bibfnamefont {M.}~\bibnamefont
  {Burrello}},\ }\bibfield  {title} {\enquote {\bibinfo {title} {Formulation of
  lattice gauge theories for quantum simulations},}\ }\href {\doibase
  10.1103/PhysRevD.91.054506} {\bibfield  {journal} {\bibinfo  {journal} {Phys.
  Rev. D}\ }\textbf {\bibinfo {volume} {91}},\ \bibinfo {pages} {054506}
  (\bibinfo {year} {2015})}\BibitemShut {NoStop}%
\bibitem [{\citenamefont {Hauke}\ \emph {et~al.}(2013)\citenamefont {Hauke},
  \citenamefont {Marcos}, \citenamefont {Dalmonte},\ and\ \citenamefont
  {Zoller}}]{hauke2013}%
  \BibitemOpen
  \bibfield  {author} {\bibinfo {author} {\bibfnamefont {P.}~\bibnamefont
  {Hauke}}, \bibinfo {author} {\bibfnamefont {D.}~\bibnamefont {Marcos}},
  \bibinfo {author} {\bibfnamefont {M.}~\bibnamefont {Dalmonte}}, \ and\
  \bibinfo {author} {\bibfnamefont {P.}~\bibnamefont {Zoller}},\ }\bibfield
  {title} {\enquote {\bibinfo {title} {Quantum simulation of a lattice
  {Schwinger} model in a chain of trapped ions},}\ }\href {\doibase
  10.1103/PhysRevX.3.041018} {\bibfield  {journal} {\bibinfo  {journal} {Phys.
  Rev. X}\ }\textbf {\bibinfo {volume} {3}},\ \bibinfo {pages} {041018}
  (\bibinfo {year} {2013})}\BibitemShut {NoStop}%
\bibitem [{\citenamefont {K\"uhn}\ \emph {et~al.}(2014)\citenamefont {K\"uhn},
  \citenamefont {Cirac},\ and\ \citenamefont {Ba\~nuls}}]{kuhn2014}%
  \BibitemOpen
  \bibfield  {author} {\bibinfo {author} {\bibfnamefont {S.}~\bibnamefont
  {K\"uhn}}, \bibinfo {author} {\bibfnamefont {J.~I.}\ \bibnamefont {Cirac}}, \
  and\ \bibinfo {author} {\bibfnamefont {M.-C.}\ \bibnamefont {Ba\~nuls}},\
  }\bibfield  {title} {\enquote {\bibinfo {title} {Quantum simulation of the
  {Schwinger} model: A study of feasibility},}\ }\href {\doibase
  10.1103/PhysRevA.90.042305} {\bibfield  {journal} {\bibinfo  {journal} {Phys.
  Rev. A}\ }\textbf {\bibinfo {volume} {90}},\ \bibinfo {pages} {042305}
  (\bibinfo {year} {2014})}\BibitemShut {NoStop}%
\bibitem [{\citenamefont {Hamer}\ \emph {et~al.}(1997)\citenamefont {Hamer},
  \citenamefont {Zheng},\ and\ \citenamefont {Oitmaa}}]{hamer1997}%
  \BibitemOpen
  \bibfield  {author} {\bibinfo {author} {\bibfnamefont {C.~J.}\ \bibnamefont
  {Hamer}}, \bibinfo {author} {\bibfnamefont {W.}~\bibnamefont {Zheng}}, \ and\
  \bibinfo {author} {\bibfnamefont {J.}~\bibnamefont {Oitmaa}},\ }\bibfield
  {title} {\enquote {\bibinfo {title} {{Series expansions for the massive
  Schwinger model in Hamiltonian lattice theory}},}\ }\href {\doibase
  10.1103/PhysRevD.56.55} {\bibfield  {journal} {\bibinfo  {journal} {Phys.
  Rev. D}\ }\textbf {\bibinfo {volume} {56}},\ \bibinfo {pages} {55--67}
  (\bibinfo {year} {1997})}\BibitemShut {NoStop}%
\bibitem [{\citenamefont {Ba\~{n}uls}\ \emph {et~al.}(2013)\citenamefont
  {Ba\~{n}uls}, \citenamefont {Cichy}, \citenamefont {Jansen},\ and\
  \citenamefont {Cirac}}]{Banuls2013}%
  \BibitemOpen
  \bibfield  {author} {\bibinfo {author} {\bibfnamefont {M.-C.}\ \bibnamefont
  {Ba\~{n}uls}}, \bibinfo {author} {\bibfnamefont {K.}~\bibnamefont {Cichy}},
  \bibinfo {author} {\bibfnamefont {K.}~\bibnamefont {Jansen}}, \ and\ \bibinfo
  {author} {\bibfnamefont {J.~I.}\ \bibnamefont {Cirac}},\ }\bibfield  {title}
  {\enquote {\bibinfo {title} {The mass spectrum of the {Schwinger} model with
  matrix product states},}\ }\href {\doibase 10.1007/JHEP11(2013)158}
  {\bibfield  {journal} {\bibinfo  {journal} {J. High Energy Phys.}\ }\textbf
  {\bibinfo {volume} {2013}},\ \bibinfo {pages} {158} (\bibinfo {year}
  {2013})}\BibitemShut {NoStop}%
\bibitem [{\citenamefont {Meurice}(2020)}]{PhysRevD.102.014506}%
  \BibitemOpen
  \bibfield  {author} {\bibinfo {author} {\bibfnamefont {Yannick}\ \bibnamefont
  {Meurice}},\ }\bibfield  {title} {\enquote {\bibinfo {title} {Discrete
  aspects of continuous symmetries in the tensorial formulation of abelian
  gauge theories},}\ }\href {\doibase 10.1103/PhysRevD.102.014506} {\bibfield
  {journal} {\bibinfo  {journal} {Phys. Rev. D}\ }\textbf {\bibinfo {volume}
  {102}},\ \bibinfo {pages} {014506} (\bibinfo {year} {2020})}\BibitemShut
  {NoStop}%
\bibitem [{\citenamefont {Unmuth-Yockey}(2019)}]{PhysRevD.99.074502}%
  \BibitemOpen
  \bibfield  {author} {\bibinfo {author} {\bibfnamefont {Judah~F.}\
  \bibnamefont {Unmuth-Yockey}},\ }\bibfield  {title} {\enquote {\bibinfo
  {title} {Gauge-invariant rotor hamiltonian from dual variables of 3d
  $u\mathbf{(}1\mathbf{)}$ gauge theory},}\ }\href {\doibase
  10.1103/PhysRevD.99.074502} {\bibfield  {journal} {\bibinfo  {journal} {Phys.
  Rev. D}\ }\textbf {\bibinfo {volume} {99}},\ \bibinfo {pages} {074502}
  (\bibinfo {year} {2019})}\BibitemShut {NoStop}%
\bibitem [{\citenamefont {Gattringer}\ \emph {et~al.}(2015)\citenamefont
  {Gattringer}, \citenamefont {Kloiber},\ and\ \citenamefont
  {M\"uller-Preussker}}]{gattringer2015}%
  \BibitemOpen
  \bibfield  {author} {\bibinfo {author} {\bibfnamefont {C.}~\bibnamefont
  {Gattringer}}, \bibinfo {author} {\bibfnamefont {T.}~\bibnamefont {Kloiber}},
  \ and\ \bibinfo {author} {\bibfnamefont {M.}~\bibnamefont
  {M\"uller-Preussker}},\ }\bibfield  {title} {\enquote {\bibinfo {title} {Dual
  simulation of the two-dimensional lattice {U(1)} gauge-{Higgs} model with a
  topological term},}\ }\href {\doibase 10.1103/PhysRevD.92.114508} {\bibfield
  {journal} {\bibinfo  {journal} {Phys. Rev. D}\ }\textbf {\bibinfo {volume}
  {92}},\ \bibinfo {pages} {114508} (\bibinfo {year} {2015})}\BibitemShut
  {NoStop}%
\bibitem [{\citenamefont {Ferguson}\ \emph {et~al.}(2021)\citenamefont
  {Ferguson}, \citenamefont {Zhang}, \citenamefont {K\"uhn}, \citenamefont
  {Wilson}, \citenamefont {Jansen},\ and\ \citenamefont {Muschik}}]{higgs1d}%
  \BibitemOpen
  \bibfield  {author} {\bibinfo {author} {\bibfnamefont {R.}~\bibnamefont
  {Ferguson}}, \bibinfo {author} {\bibfnamefont {J.}~\bibnamefont {Zhang}},
  \bibinfo {author} {\bibfnamefont {S.}~\bibnamefont {K\"uhn}}, \bibinfo
  {author} {\bibfnamefont {C.~M.}\ \bibnamefont {Wilson}}, \bibinfo {author}
  {\bibfnamefont {K.}~\bibnamefont {Jansen}}, \ and\ \bibinfo {author}
  {\bibfnamefont {C.~A.}\ \bibnamefont {Muschik}},\ }\href@noop {} {\bibfield
  {journal} {\bibinfo  {journal} {Manuscript in preparation}\ } (\bibinfo
  {year} {2021})}\BibitemShut {NoStop}%
\bibitem [{\citenamefont {Bhanot}\ and\ \citenamefont
  {Creutz}(1980)}]{bhanot1980phase}%
  \BibitemOpen
  \bibfield  {author} {\bibinfo {author} {\bibfnamefont {G.}~\bibnamefont
  {Bhanot}}\ and\ \bibinfo {author} {\bibfnamefont {M.}~\bibnamefont
  {Creutz}},\ }\bibfield  {title} {\enquote {\bibinfo {title} {{Phase diagram
  of $Z(N)$ and U(1) gauge theories in three dimensions}},}\ }\href {\doibase
  10.1103/PhysRevD.21.2892} {\bibfield  {journal} {\bibinfo  {journal} {Phys.
  Rev. D}\ }\textbf {\bibinfo {volume} {21}},\ \bibinfo {pages} {2892}
  (\bibinfo {year} {1980})}\BibitemShut {NoStop}%
\bibitem [{\citenamefont {Jers{\'a}k}\ \emph {et~al.}(1983)\citenamefont
  {Jers{\'a}k}, \citenamefont {Neuhaus},\ and\ \citenamefont
  {Zerwas}}]{jersak1983u}%
  \BibitemOpen
  \bibfield  {author} {\bibinfo {author} {\bibfnamefont {J.}~\bibnamefont
  {Jers{\'a}k}}, \bibinfo {author} {\bibfnamefont {T.}~\bibnamefont {Neuhaus}},
  \ and\ \bibinfo {author} {\bibfnamefont {P.~M.}\ \bibnamefont {Zerwas}},\
  }\bibfield  {title} {\enquote {\bibinfo {title} {U(1) lattice gauge theory
  near the phase transition},}\ }\href {\doibase 10.1016/0370-2693(83)90115-6}
  {\bibfield  {journal} {\bibinfo  {journal} {Phys. Lett. B}\ }\textbf
  {\bibinfo {volume} {133}},\ \bibinfo {pages} {103} (\bibinfo {year}
  {1983})}\BibitemShut {NoStop}%
\bibitem [{\citenamefont {Bhanot}(1982)}]{bhanot1982compact}%
  \BibitemOpen
  \bibfield  {author} {\bibinfo {author} {\bibfnamefont {G.}~\bibnamefont
  {Bhanot}},\ }\bibfield  {title} {\enquote {\bibinfo {title} {Compact {QED}
  with an extended lattice action},}\ }\href {\doibase
  10.1016/0550-3213(82)90382-0} {\bibfield  {journal} {\bibinfo  {journal}
  {Nucl. Phys. B}\ }\textbf {\bibinfo {volume} {205}},\ \bibinfo {pages} {168}
  (\bibinfo {year} {1982})}\BibitemShut {NoStop}%
\bibitem [{\citenamefont {Adler}(1969)}]{Adler1969}%
  \BibitemOpen
  \bibfield  {author} {\bibinfo {author} {\bibfnamefont {S.~L.}\ \bibnamefont
  {Adler}},\ }\bibfield  {title} {\enquote {\bibinfo {title} {{Axial-Vector
  Vertex in Spinor Electrodynamics}},}\ }\href {\doibase
  10.1103/PhysRev.177.2426} {\bibfield  {journal} {\bibinfo  {journal} {Phys.
  Rev.}\ }\textbf {\bibinfo {volume} {177}},\ \bibinfo {pages} {2426} (\bibinfo
  {year} {1969})}\BibitemShut {NoStop}%
\bibitem [{\citenamefont {Bell}\ and\ \citenamefont {Jackiw}(1969)}]{Bell1969}%
  \BibitemOpen
  \bibfield  {author} {\bibinfo {author} {\bibfnamefont {J.~S.}\ \bibnamefont
  {Bell}}\ and\ \bibinfo {author} {\bibfnamefont {R.}~\bibnamefont {Jackiw}},\
  }\bibfield  {title} {\enquote {\bibinfo {title} {{A PCAC puzzle:
  $\pi_0\rightarrow\gamma\gamma$ in the $\sigma$-model}},}\ }\href {\doibase
  10.1007/BF02823296} {\bibfield  {journal} {\bibinfo  {journal} {Nuovo Cimento
  A}\ }\textbf {\bibinfo {volume} {60}},\ \bibinfo {pages} {47} (\bibinfo
  {year} {1969})}\BibitemShut {NoStop}%
\bibitem [{\citenamefont {Fujikawa}(1979)}]{Fujikawa1979}%
  \BibitemOpen
  \bibfield  {author} {\bibinfo {author} {\bibfnamefont {K.}~\bibnamefont
  {Fujikawa}},\ }\bibfield  {title} {\enquote {\bibinfo {title} {Path-integral
  measure for gauge-invariant fermion theories},}\ }\href {\doibase
  10.1103/PhysRevLett.42.1195} {\bibfield  {journal} {\bibinfo  {journal}
  {Phys. Rev. Lett.}\ }\textbf {\bibinfo {volume} {42}},\ \bibinfo {pages}
  {1195} (\bibinfo {year} {1979})}\BibitemShut {NoStop}%
\bibitem [{\citenamefont {Dunne}\ and\ \citenamefont
  {Trugenberger}(1989)}]{Dunne1989}%
  \BibitemOpen
  \bibfield  {author} {\bibinfo {author} {\bibfnamefont {G.~V.}\ \bibnamefont
  {Dunne}}\ and\ \bibinfo {author} {\bibfnamefont {C.~A.}\ \bibnamefont
  {Trugenberger}},\ }\bibfield  {title} {\enquote {\bibinfo {title} {{Kinetic
  normal ordering and the Hamiltonian structure of $U(1)$ chiral anomalies in
  3+1 dimensions}},}\ }\href {\doibase 10.1016/0003-4916(89)90248-0} {\bibfield
   {journal} {\bibinfo  {journal} {Ann. Phys.}\ }\textbf {\bibinfo {volume}
  {195}},\ \bibinfo {pages} {356} (\bibinfo {year} {1989})}\BibitemShut
  {NoStop}%
\end{thebibliography}%
\end{document}